\documentclass[11pt,a4paper]{article}
\usepackage{epsfig,axodraw}
\usepackage{array,cite,amsmath,amssymb,multirow}
\usepackage[flushleft]{threeparttable}
\usepackage{subcaption}
\usepackage{kantlipsum}
\usepackage[usenames,dvipsnames]{xcolor}
\usepackage[breakable, theorems, skins]{tcolorbox}
\tcbset{enhanced}
\DeclareRobustCommand{\mybox}[2][gray!20]{%
\begin{tcolorbox}[   
        breakable,
        left=0pt,
        right=0pt,
        top=0pt,
        bottom=0pt,
        colback=#1,
        colframe=#1,
        width=\dimexpr\textwidth\relax, 
        enlarge left by=0mm,
        boxsep=5pt,
        arc=0pt,outer arc=0pt,
        ]
        #2
\end{tcolorbox}
}
\usepackage{lineno}
\usepackage{tikz}
\usepackage[compat=1.1.0]{tikz-feynman}
\usepackage{listings}
\usepackage{hhline}
\lstdefinestyle{mystyle}{
    backgroundcolor=\color{lightgray}
}
\newcommand{\newc}{\newcommand}
\newc{\gev}{\,GeV}
\newcolumntype{M}[1]{>{\centering\arraybackslash}m{#1}}
\newcolumntype{N}{@{}m{0pt}@{}}
\newc{\mev}{\,MeV}
\newc{\ra}{\rightarrow}
\newc{\rpv}{$\mathrm{\not\!R_p}$}
\newc{\rp}{$\mathrm{R_p}$}
\newc{\real}{\mathcal{R}e}
\newc{\alsm}{{\displaystyle \sum_{\alpha=1,2}}}
\newc{\besm}{{\displaystyle \sum_{\beta=1,2}}}
\newc{\al}{\alpha}
\newc{\sgn}{\mr{sgn}\,}
\newc{\be}{\beta}
\newc{\ga}{\gamma}
\newc{\de}{\delta}
\newc{\sla}{\!\!\!\!\!\not\:\:\!}
\newc{\slab}{\!\!\!\!\!\not\,\,\,}
\newc{\slac}{\!\!\!\!\!\!\!\not\,\,\,\,}
\newc{\met}{$\not\!\!E_T$}
\newc{\cw}{\cos\theta_W}
\newc{\sw}{\sin\theta_W}
\newc{\ssw}{\sin^2\theta_W}
\newc{\ccw}{\cos^2\theta_W}
\newc{\cbe}{\cos\beta}
\newc{\sbe}{\sin\beta}
\newc{\ort}{\frac1{\sqrt{2}}}
\newc{\sh}{\hat{s}}
\newc{\uh}{\hat{u}}
\newc{\tha}{\hat{t}}
\newc{\sa}{\sin\al}
\newc{\ca}{\cos\al}
\newc{\mz}{M_{\mr{Z}}}
\newc{\mw}{M_{\mr{W}}}
\newc{\bv}{$\mathrm{\not\!B}$}
\newc{\lv}{$\mathrm{\not\!L}$}
\newc{\beq}{\begin{equation}}
\newc{\eeq}{\end{equation}}
\newc{\ie}{{\it i.e.\/}\ }
\newc{\lam}{\lambda}
\newc{\cht}{\tilde{\chi}}
\newc{\glt}{\tilde{g}}
\newc{\upt}{\tilde{u}}
\newc{\qkt}{\tilde{q}}
\newc{\elt}{\tilde{\ell}}
\newc{\hgt}{\tilde{H}}
\newc{\nut}{\tilde{\nu}}
\newc{\dnt}{\tilde{d}}
\newc{\ftl}{\mr{\tilde{f}}}
\newc{\psb}{\bar{\psi}}
\newc{\rtt}{2^{1/2}}
\newc{\mut}{\tilde{\mu}}
\newc{\mr}{\mathrm}
\newc{\bath}{\bar{\theta}}
\newc{\tht}{\theta}
\newc{\JC}{{\bf J}}
\newc{\lra}{\longrightarrow}
\newc{\eg}{{\it e.g.\  }}
\newc{\barr}{\begin{eqnarray}}
\newc{\earr}{\end{eqnarray}}
\newc{\me}{\mathcal{M}}
\newc{\dbm}{\partial_\mu}
\newc{\dbmu}{\stackrel{\leftrightarrow\  }{\partial^\mu}}
\newc{\sgm}{\sigma_\mu}
\newc{\captionB}[2]{\caption[{#1}]{{\small {#2}}}}
\newc{\ahref}[2]{#2}
\usepackage{jheppub} 

\title{Generalized angular-order parton showers in \textsf{Herwig~7}}

\author[a]{Joon-Bin Lee}
\author[b]{M.R. Masouminia}
\author[c]{Michael H. Seymour}
\author[a]{Un-ki Yang}

\affiliation[a]{Department of Physics and Astronomy, Seoul National University,\\Seoul, Republic of Korea}
\affiliation[b]{Institute for Particle Physics Phenomenology, Durham University,\\Durham, U.K.}
\affiliation[c]{Department of Physics and Astronomy, University of Manchester,\\Manchester, U.K.}

\emailAdd{joon.bin.lee@cern.ch}
\emailAdd{mohammad.r.masouminia@durham.ac.uk}
\emailAdd{michael.seymour@manchester.ac.uk}
\emailAdd{ukyang@snu.ac.kr}

\abstract{
This paper presents the inaugural investigation of beyond the Standard Model (BSM) radiation processes, framed as a generalized, process- and model-independent parton shower algorithm within \textsf{Herwig~7}, based on direct translations of Universal FeynRules Output (UFO) constructed via \textsf{Herwig}'s \texttt{ufo2herwig} module. Leveraging the fact that shower kinematics are dictated by the spins of involved particles, we calculate comprehensive helicity-dependent branching kernels for all feasible splittings of scalars, fermions, and vector bosons, tailored to \textsf{Herwig~7}'s angular-ordering (AO) parton shower algorithm. Utilizing these kernels, we derive BSM splitting functions in the quasi-collinear limit, ensuring compatibility with the Standard Model (SM) and supersymmetry (SUSY) splitting functions when analogous parameter conditions are applied. These newly derived functions have been integrated into the \textsf{Herwig~7} event generator framework. Comparative analyses with fixed-order matrix element calculations show good agreement for single radiation events. Moreover, the results showcase the influence of BSM radiation at the Large Hadron Collider (LHC) and envisage its implications for future collider endeavours. This research augments our comprehension of BSM radiation effects, with significant bearings on present and prospective collider-based inquiries.
}

\begin{document}
\noindent{\hfill \small MCnet-23-21}

\noindent{\hfill \small IPPP/23/73\\[0.1in]}
\maketitle
\flushbottom

\section{Introduction}

The discovery of the Higgs boson at the Large Hadron Collider (LHC) was a significant event, confirming the Standard Model (SM) of particle physics. Yet, it did not provide answers to all our questions. Mysteries such as the constituents of dark matter, the small but non-zero neutrino masses~\cite{nu_oscillation_theory, nu_oscillation_kamiokande, nu_oscillation_SNO}, the imbalance between matter and antimatter, the recently observed discrepancies in the W boson's mass~\cite{w_mass_problem-cdf}, B-physics anomalies~\cite{b_anomaly_summary}, the muon $g-2$ anomaly from Fermilab~\cite{muon_g-2_experiment}, and the enigmatic hierarchy problem remain unresolved. These open questions provide a strong motivation for theories beyond the Standard Model (BSM). The LHC, along with future collider projects, is poised to explore these mysteries by venturing into uncharted energy territories where new physics -- including supersymmetry (SUSY), extra spatial dimensions, and composite Higgs models -- may be discovered.

Searches for alternative Higgs bosons other than the observed 125 GeV particle, as well as for hypothetical $Z'$ and $W'$ bosons, are central to probing BSM phenomena, each characterized by unique coupling patterns not seen in the SM. For simulating such collider events, multi-purpose event generators like \textsf{Herwig~7}~\cite{herwig_manual, herwig7.0_release_note, herwig7.2_release_note, Bewick:2023tfi, Masouminia:2023zhb}, \textsf{PYTHIA 8}~\cite{pythia_manual}, and \textsf{Sherpa}~\cite{Sherpa1.1_manual, Sherpa2.2_manual, Sherpa_BSM} are indispensable tools. \textsf{Sherpa} often operates independently to integrate both the calculation of the hard process and the subsequent parton shower, whereas \textsf{PYTHIA 8} and \textsf{Herwig~7} utilize auxiliary programs like \textsf{MadGraph5}~\cite{MG5_manual, MG5_BSM} to generate BSM hard processes and concentrate on an accurate description of the parton showering and hadronization of those processes. These computational frameworks have been validated through their success in accurately reproducing experimental data and in their capability to handle both SM and BSM scenarios, incorporating perturbative QCD computations as well.

Nevertheless, at the heightened centre-of-mass energies probed by the LHC, the production of massive weakly interacting bosons often features collinear and soft kinematics, necessitating the inclusion of massive boson radiation in data analyses~\cite{neccessity_of_ewk_radiation, Darvishi:2020paz}. A notable advancement in general-purpose event generators is the refined capability to simulate massive boson radiation accurately~\cite{pythia_ewk_radiation, herwig_ewk_radiation, vincia_ewk_radiation, ewk_radiation-theory_GET}. Leveraging this capability, it is now feasible to model the radiative production of BSM bosons. However, adapting current parton shower algorithms to accommodate BSM bosons poses a challenge due to their unique kinematic properties compared to their SM counterparts. This includes the need to account for CP-odd couplings associated with non-SM Higgs and vector bosons, which interact through right-handed couplings absent in the SM framework.

The LHC collaborations have recently intensified their search for low-mass BSM particles~\cite{CMS-low_mass_Z'-1, CMS-low_mass_Z'-2, CMS-low_mass_Z'-3, LHCb-low_mass_Z'}. This shift in focus is primarily because the last decade did not yield direct evidence of new particles at higher mass scales. For low-mass particles, the likelihood of their production through radiative processes is relatively higher than for their high-mass counterparts. Accurate simulation of such radiative production is crucial; otherwise, it could significantly alter the kinematic features expected from matrix element calculations. Consequently, this necessitates the development of new, more general splitting functions to properly model the BSM parton showers. 

As well as the theoretical advantage of implementing BSM showers to simulate multi-emission processes, there is also an important practical advantage. The shower algorithms already work through all final-state particles considering whether to generate QCD, QED or EW radiation from them. The additional time taken by adding BSM radiation is small and grows only slowly with the number of emitted particles. On the other hand, the time needed for fixed order calculations grows at least factorially with the number of produced particles so, at high energies, only the showering approach is viable. This speed advantage is particularly important because, when probing BSM model space, it is usually necessary to make multiple runs over many different parameter value sets. Moreover, the angular-ordered (AO) parton shower in \textsf{Herwig~7}, which implements intrajet and interjet effects due to colour coherence~\cite{intro_to_AO}, has recently been adapted to keep track of the opening angle conditions for EW radiation~\cite{herwig_ewk_radiation} making it ready also for BSM radiation. The main missing ingredients are the set of splitting functions needed and the important infrastructure to import the model data needed to initialise these.

In this paper, we present generalized spin-unaveraged splitting functions for the AO parton shower algorithm applicable to fermions, scalar bosons, and vector bosons. These functions are model-independent and are designed to cover a broad spectrum of particle splittings, including $\phi \rightarrow \phi'\phi''$, $f\rightarrow f'\phi$, $V \rightarrow V'\phi$, $\phi\rightarrow \phi'V$, $V\rightarrow \phi\phi'$, $f\rightarrow f'V$, and $V\rightarrow V'V''$ transitions. The advantage of spin-unaveraged splitting functions lies in their ability to incorporate the spin-dependence of interactions, which is particularly critical in BSM physics where couplings between quarks and bosons are spin-sensitive. For instance, the interaction of a scalar boson with a fermion pair is characterized by:
\begin{equation}
\begin{split}
&V_{qq'H} = -ig\kappa \quad \;\;\; \text{(for a CP-even scalar)} \\
&V_{qq'H} = -ig\Tilde{\kappa} \gamma_5 \quad \text{(for a CP-odd scalar)}
\end{split}
\label{eq:V_ffs}
\end{equation}
Note that the Standard Model does not include a CP-odd Higgs boson. For vector bosons interacting with fermions, the Standard Model prescribes specific left- and right-handed couplings, represented by:
\begin{equation}
V_{qq'V} = -ig(g_LP_L+g_RP_R)\gamma^\mu.
\end{equation}
To accurately simulate these interactions, modifications to the internal mechanisms of event generators like \textsf{Herwig~7} are necessary. These changes allow the generators to interpret values from Universal FeynRules Output (UFO) models~\cite{feynrules2_manual, UFO_model_file} and correctly apply them to the splittings. In BSM scenarios, where complex couplings often occur, spin-unaveraged splitting functions are indispensable not only for correctly handling polarization effects but also for managing the influence of heavy particles, where spin orientations significantly affect the kinematic distributions of final-state particles.

The Angular-Ordered Parton Shower algorithm in \textsf{Herwig~7}~\cite{herwig_manual, herwig7.0_release_note, Bewick:2019rbu, herwig_ewk_radiation, Bewick:2021nhc} and transverse-momentum ordered parton showers, such as \textsf{DIRE}~\cite{Hoche:2015sya}, along with those in \textsf{PanScales}~\cite{Dasgupta:2020fwr} and \textsf{Alaric}~\cite{Nagy:2020rmk}, differ fundamentally in their ordering mechanisms, which significantly influences the phenomenological outcomes of the simulations. Despite these differences, the calculation of quasi-collinear, helicity-dependent splitting functions is based on universal principles, chiefly the spins of the interacting particles and the Feynman rules governing their interactions. This foundational aspect implies that, aside from minor convention adjustments, these splitting functions are transferable across different parton shower models. This transferability allows for a unified approach in improving the accuracy of simulations, though the incorporation of these functions into a specific parton shower algorithm may still need to account for the algorithm's unique features and approximations.

Furthermore, although we accurately calculate these spin-unaveraged splitting functions, there are challenges when simulating massive particle splittings. Specifically, longitudinally polarized vector bosons can lead to divergences as their mass approaches zero, which poses numerical difficulties in event generators. To circumvent this, we adopt Dawson's method~\cite{Dawson}, which mitigates the issue by eliminating terms that are inversely proportional to the boson mass, $m_i$. Another nuanced problem arises from the inapplicability of the collinear limit of massless splittings to massive ones. However, the quasi-collinear approach can be utilized to achieve a comparable suppression in the forward region, analogous to the collinear limits~\cite{herwig_manual, SUSY_splitting-2}. The resultant splitting functions bear similarities to those in the Standard Model~\cite{Altarelli_PS_introduction, 2_loop_sf, mhv_rules}, suggesting their credibility even prior to detailed verification. Ultimately, these generalized splitting functions are integrated into the \textsf{Herwig~7} event generator, enabling BSM simulations. Validation of these results, alongside comparative studies with \textsf{MadGraph5} leading-order computations, was conducted across diverse model configurations, including the general two-Higgs-doublet model (2HDM)~\cite{2hdm-ufo,Darvishi:2019ltl,Darvishi:2023nft}, the minimal B-L extension of the SM~\cite{bl4-ufo-1, bl4-ufo-2, bl4-ufo-3}, and the $W'$ effective model~\cite{effW-ufo-1, effW-ufo-2}. These models were provided in the UFO format, facilitating their integration and testing within the event generator framework.

The structure of this paper is outlined as follows: Section~\ref{sec:kinematics} details the essential components for computing quasi-collinear, spin-dependent splitting functions, encompassing the kinematics and dynamics of partons in the quasi-collinear limit. We elucidate the steps necessary to derive these splitting functions in a universal form, suitable for integration with the \textsf{Herwig~7} event generator. In subsequent sections, we address the emissions of particles with various spins. Specifically, section~\ref{sec:higgs} discusses the emission of spin-0 particles from parents of spin-0, -1/2, and -1, while section~\ref{sec:vector} explores the emission of spin-1 particles, covering all relevant electroweak (EW) BSM couplings for both massive and massless partons. In section~\ref{sec:result}, we present a comparative analysis between fixed-order \textsf{MadGraph5} simulations of $n+1$ final state particle processes and the resummed results of showers for $n$ final state particle hard processes from \textsf{MadGraph5} with an accompanying \textsf{Herwig~7} shower that permits only a single radiation event, thus emulating the $n+1$ particle process of \textsf{MadGraph5}. This comparison allows us to evaluate the effects of a full \textsf{Herwig~7} SM+BSM shower against the \textsf{MadGraph5} matrix element calculations that only incorporate SM showers (both cases include interleaved QCD+QED+EW AO showers). The concluding section, Section~\ref{sec:conclusion}, encapsulates our findings, with a particular focus on the model-independent utility of the generalized splitting functions. Further practical guidance on executing BSM showers within the \textsf{Herwig~7} framework is furnished in Section~\ref{sec:manual}.

\section{Parton shower kinematics in the quasi-collinear limit}
\label{sec:kinematics}

In this section, we provide an overview of the essential components required to compute generalized angular-ordered splitting functions within the quasi-collinear limit~\cite{heavyquark, SUSY_splitting-2, herwig_ewk_radiation}. Considering a generic splitting $0 \to 1,2$, an AO parton shower can be characterized by the light-cone momentum fraction, $z$, which parameterizes the momentum component of the parent particle in the direction of the child particle, and the evolution scale, $\tilde{q}$. The evolution scale by default is defined as
\begin{align}
\tilde{q}^2  = { 2q_1 \cdot q_2 + m_1^2 + m_2^2 - m_0^2 \over z (1-z) },
\label{dot}
\end{align}
within \textsf{Herwig}'s so-called ``dot-product-preserving" scheme~\cite{Bewick:2019rbu, Bewick:2021nhc}, although other choices are also available, i.e. the $p_T$-preserving~\cite{Gieseke:2003rz, Gieseke:2003hm} and $q^2$-preserving~\cite{Reichelt:2017hts, herwig7.1_release_note} schemes. In Eq.~\eqref{dot}, the momenta of the participating partons are defined using the Sudakov decomposition as
\begin{align}
    q_i = \alpha_i p + \beta_i n + \gamma_i q_T,
\label{eq:qi}
\end{align}
where $p$ is a four-vector of the on-shell progenitor before branching, $n$ is a reference vector, and $q_T$ is the transverse part of the particle momentum. If we choose the $z$ axis in the direction of the incoming parton, we obtain $p = (\sqrt{\textbf{p}^2 + m_0^2}, 0, 0, \textbf{p})$ and $n = (1, 0, 0, -1)$. The Sudakov parameters $\alpha_i$, $\beta_i$ and $\gamma_i$ are defined as
\begingroup
\vspace{-0.25in}
\begin{flushleft}
\begin{minipage}[t]{.25\columnwidth}
    \begin{align*}
        \alpha_i = 
        \begin{Bmatrix}
            &1   \\
            &z   \\
            &1-z 
        \end{Bmatrix},
    \end{align*}
\end{minipage}
\begin{minipage}[t]{.3\columnwidth}
    \begin{align*}
        \beta_i = 
        \begin{Bmatrix}
            &\beta_1+\beta_2 \\
            &\frac{p_T^2+m_1^2-z^2 m_0^2}{2zp\cdot n} \\
            &\frac{p_T^2+m_2^2-(1-z)^2m_0^2}{2(1-z)p\cdot n}
        \end{Bmatrix},
    \end{align*}
\end{minipage}
\begin{minipage}[t]{.4\columnwidth}
    \begin{align}
        \gamma_i = 
        \begin{Bmatrix}
            &0\\
            &1\\
            &-1
        \end{Bmatrix},
        \quad\text{for}\;i=
        \begin{Bmatrix}
            &0& \\
            &1& \\
            &2&
        \end{Bmatrix}.
    \end{align}
\end{minipage}
\end{flushleft}
\endgroup

In attempting to calculate the splitting function for the generic $0 \to 1,2$ splitting, one needs to define the spinors and polarization vectors of the progenitor and the children. The spinors are conventionally defined in the Chiral basis as
\begin{subequations}
\begin{align}
    u_{\frac{1}{2}}(p) &= \left[ \frac{m_0}{\sqrt{2p}}; 0; \sqrt{2p}\left( 1+\frac{m_0^2}{8p^2} \right); 0 \right],
\\
    u_{-\frac{1}{2}}(p) &= \left[ 0; \sqrt{2p}\left( 1+\frac{m_0^2}{8p^2} \right); 0; \frac{m_0}{\sqrt{2p}} \right],
\\
    u_{\frac{1}{2}}(q_1) &= \left[ \frac{m_1}{\sqrt{2zp}}; \frac{e^{j\phi}m_1p_T}{(2zp)^{3/2}}; \sqrt{2zp}\left( 1+\frac{m_0^2}{8p^2} \right); \frac{e^{j\phi}p_T}{\sqrt{2zp}} \right],
\\
    u_{-\frac{1}{2}}(q_1) &= \left[ -\frac{e^{-j\phi}p_T}{\sqrt{2zp}}; \sqrt{2zp}\left( 1+\frac{m_0^2}{8p^2} \right); -\frac{e^{-j\phi}m_1p_T}{(2zp)^{3/2}}; \frac{m_1}{\sqrt{2zp}} \right],
\end{align}
\end{subequations}
with $\Bar{u}_{\pm\frac{1}{2}}(q_i) = u_{\pm\frac{1}{2}}^{\dag}(q_i) \gamma^0$ and $j$ denotes $\sqrt{-1}$. Note that within \textsf{Herwig}'s parton shower algorithm, a splitting is boosted to a reference frame where the progenitor propagates along the $z$ axis, without loss of generality. After determining the kinematics of the splitting, the entire frame is then boosted back to the original frame, and the cascade continues.

The polarization vectors of incoming vector bosons are given as follows:
\begin{subequations}
    \begin{align}
    \epsilon_{\lambda_0 = \pm1}^{\mu}(p) &= -\frac{1}{\sqrt{2}}\left( 0; \lambda_0; j; 0\right),
	\\
    \epsilon_{\lambda_0 = 0}^{\mu}(p) &= \frac{1}{m_0}\left( p; 0; 0; \sqrt{m_0^2+p^2} \right).
\end{align}
\end{subequations}
On the other hand, the polarization vectors of the outgoing vector bosons can be derived in the following forms~\cite{herwig_ewk_radiation};
\begin{subequations}
\begin{align}
    \epsilon^\mu_{\lambda_i = \pm 1} (q_i) &= 
    \left[
    0; 
    -\frac{\lambda_i}{\sqrt{2}}\left( 1-\frac{p_T^2e^{j\lambda_i\phi}\cos\phi}{2\alpha_i^2p^2} \right); 
    -\frac{j}{\sqrt{2}}+\frac{\lambda_ip_T^2e^{j\lambda_i\phi}\sin\phi}{2\sqrt{2}\alpha_i^2p^2}; 
    \gamma_i \frac{\lambda_ip_Te^{j\lambda_i\phi}}{\sqrt{2}\alpha_i p} 
    \right]
\\
    \epsilon^\mu_{\lambda_i = 0} (q_i) &= 
    \bigg[ 
    \frac{\alpha_i p}{m_i}+\frac{p_T^2+\alpha_i^2m_0^2-m_i^2}{4\alpha_i pm_i}; 
    \gamma_i \cos\phi \left( \frac{p_T}{m_i}+\frac{m_ip_T}{2\alpha_i^2p^2} \right);
    \gamma_i \sin\phi \left( \frac{p_T}{m_i}+\frac{m_ip_T}{2\alpha_i^2p^2} \right);
    \nonumber \\ & \;\;\;\;\;\;
    \frac{\alpha_i p}{m_i} - \frac{p_T^2-\alpha_i^2m_0^2-m_i^2}{4\alpha_i pm_i} 
    \bigg], 
\end{align}
\label{eq:polarization_qi}
\end{subequations} 
when $i = 1, 2$.

One immediately notes the presence of $q_i/m_i$ terms in Eqs.~\eqref{eq:polarization_qi} that are expected to diverge in either the Breit momentum frame or for an unbroken electroweak theory. It is, however, possible to utilise Dawson's approach~\cite{Dawson} and subtract all the terms proportional to $q_i/m_i$ from the polarisation vector in order to deal with these terms in the longitudinal polarizations~\cite{herwig_ewk_radiation}. Following this disposition, the longitudinal part of a vector boson is curtailed to
\begin{equation}
    \epsilon^\mu_{\lambda_i = 0^*}(q_i) = \epsilon_{\lambda_i = 0}^{\mu}(q_i) -\frac{q_i}{m_i} = \frac{m_i}{2\alpha_i p} \left( -1; \frac{\cos\phi p_T}{\alpha_i p}; \frac{\sin\phi p_T}{\alpha_i p}; 1 \right)
\label{eq:polarization_qi*}
\end{equation}

Finally, we can calculate the splitting functions of showering processes following Altarelli and Parisi's method introduced in ref.~\cite{Altarelli_PS_introduction}. This paper introduces a master formula to compute the splitting probability\footnote{The functions $P(z,\Tilde{q})$ we introduce give the probability distributions in both $z$ and $\Tilde{q}$, so we refer to them as splitting \emph{probabilities}. They are proportional to the conventional splitting functions, which are the probability distributions in $z$ at fixed $\Tilde{q}$.} using a matrix element of a local process independent of the hard process, which is
\begin{equation}
    P_{0\rightarrow 12}(z, \Tilde{q}) = \frac{1}{2(q_0^2-m_0^2)} \sum_{s_0,s_1,s_2}|\mathcal{M}_{s_0,s_1,s_2}|^2,
\label{eq:derivation}
\end{equation}
where $s_i \; (i=0,1,2)$ denotes the spins of the particles participating in the process. We will thus calculate matrix elements for all incoming and outgoing spin combinations, insert them into the above equation, and then evaluate the splitting function for all kinds of branching processes. 

\section{Scalar particle splittings}
\label{sec:higgs}

Given the inherently simpler Feynman rules for the incoming/outgoing scalar particles, we begin the calculation of the generalized splitting functions with scalar particle emissions from various types of currents, starting from spin-0 Higgs-like currents to spin-1/2 fermion and spin-1 vector boson currents. We have calculated these splitting functions and compared the results with the SM Higgs splitting outcomes~\cite{herwig_ewk_radiation}, as well as with the branching behaviours of massless squarks and gluinos, which possess spins of $0$ and $1/2$ respectively, in the context of SUSY theory~\cite{SUSY_splitting, SUSY_splitting-2}. A notable distinction from the SM EW boson splitting functions is the potential for the incoming fermion or vector boson to change its flavour via charged scalars or flavour-changing neutral currents (FCNCs).

\subsection{\texorpdfstring{$\phi\rightarrow \phi'\phi''$}{phi to phi'phi''} splitting function}\label{sec:sss}

We begin with the $\phi \rightarrow \phi' \phi''$ splitting notably via the triple Higgs couplings. This type of splitting is a common feature not only in the SM but also in theories with additional Higgs bosons, such as 2HDM~\cite{2HDM_THC-1, 2HDM_THC-2,Darvishi:2019ltl,Darvishi:2023nft} and SUSY~\cite{SUSYhiggs,Pilaftsis:1999qt}. The vertex factor for this splitting is given by a coupling constant itself without any additional factors, $ig$, leading to the invariant matrix element:
\begin{equation}
    -i\mathcal{M} \left[ 
    \raisebox{-0.38in}{\includegraphics[clip, width=0.12\textwidth]{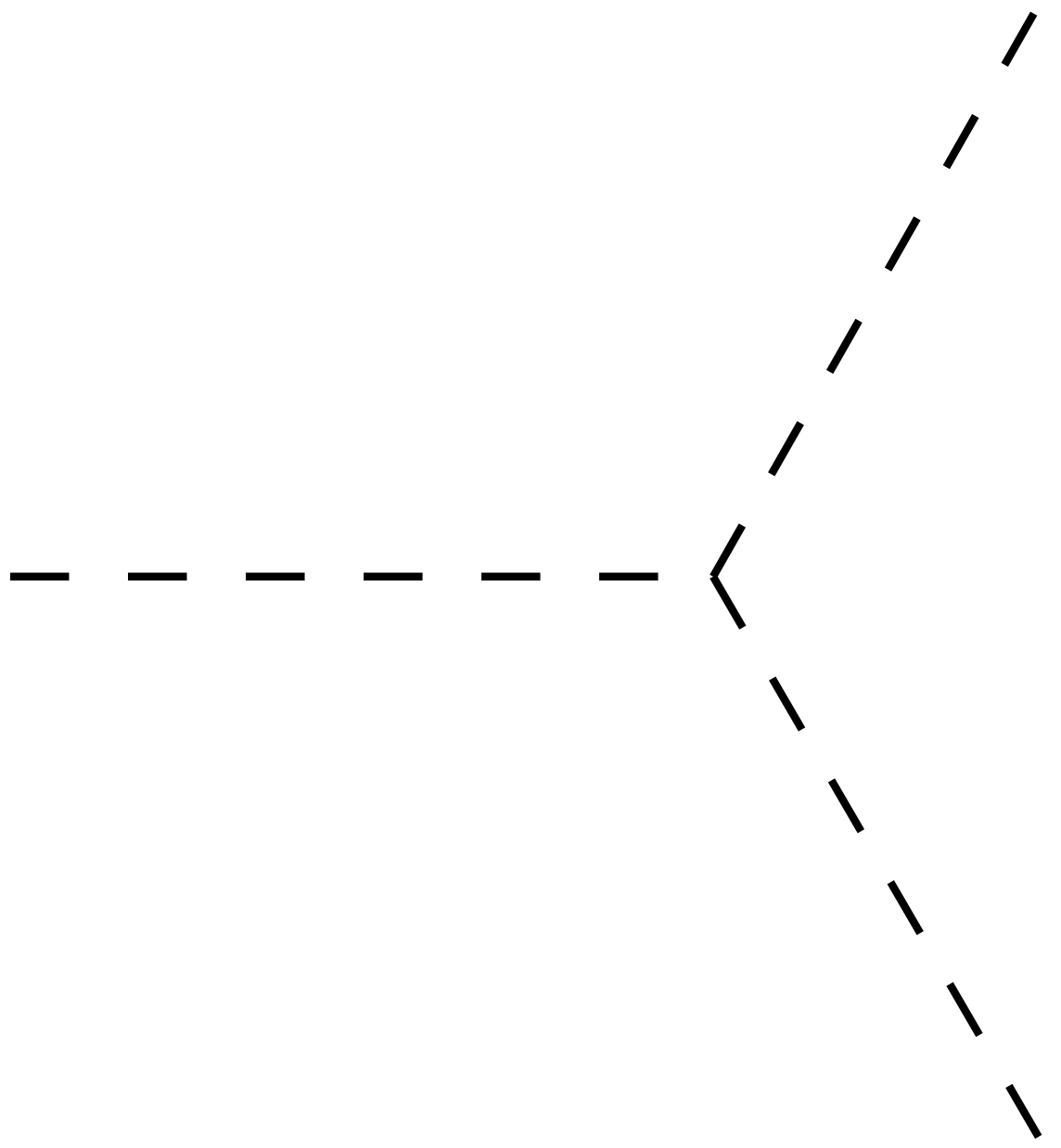}}
    \right] 
    = -ig,
\end{equation}
This invariant matrix element, in conjunction with~\eqref{eq:derivation}, yields the splitting probability for the triple scalar boson coupling as follows:
\begin{equation}
    P_{\phi \rightarrow \phi' \phi''}(z, \tilde{q}) = \frac{g^2}{2Sz(1-z)\tilde{q}^2},
\label{eq:sss splitting function}
\end{equation}
where the symmetry factor $S$ is $1$ for $\phi' \neq \phi''$ and $2$ for $\phi' = \phi''$. It is important to note that this symmetry factor becomes unity in the broken electroweak limit, where the scalar boson is produced with asymmetric kinematics, rendering the two final state partons distinguishable.

In general, the coupling is given by $g = n!v\lambda$, where the factor $n$ is a symmetry factor, $v$ represents the vacuum expectation value, and $\lambda$ is the value of the triple Higgs coupling associated with the Higgs quartic potential. Typical forms of $\lambda$ in 2HDM are detailed in Appendix A of ref.~\cite{2HDM_THC-1}. For the SM Higgs boson, where $m_h = \sqrt{2\lambda v^2}$ and $m_W = \frac{1}{2}vg_W$, the coupling is expressed as $g_{hhh}= \frac{3}{2} g_W \frac{m_h^2}{m_W}$. Consequently, the SM Higgs to di-Higgs splitting probability is formulated as:
\begin{align}
P_{h\rightarrow hh}(z, \tilde{q}) = \frac{g_W^2}{z(1-z)\tilde{q}^2}\frac{9m_h^4}{16m_W^2}.
\end{align}

\subsection{\texorpdfstring{$f\rightarrow f'\phi$}{f to f'phi} splitting function} \label{sec:ffh}

The $q \rightarrow q'\phi$ branching in BSM scenarios exhibits richer phenomenology compared to the SM Higgs boson branchings. One notable aspect is the presence of flavour-changing currents, which can occur via charged scalar bosons or FCNCs, both commonly found in many BSM theories. Another aspect is the CP-odd (pseudoscalar) coupling, characterized by an additional $\gamma_5$ term in the vertex factor, in contrast to the CP-even (scalar) coupling, whose vertex factor is simply $i\kappa$. Our analysis takes into account all these additional characteristics inherent to BSM theories.

The matrix element of a generic $q \rightarrow q'\phi$ branching is given as
\begin{equation}
    -i\mathcal{M}\left[
    \raisebox{-0.38in}{\includegraphics[clip, width=0.12\textwidth]{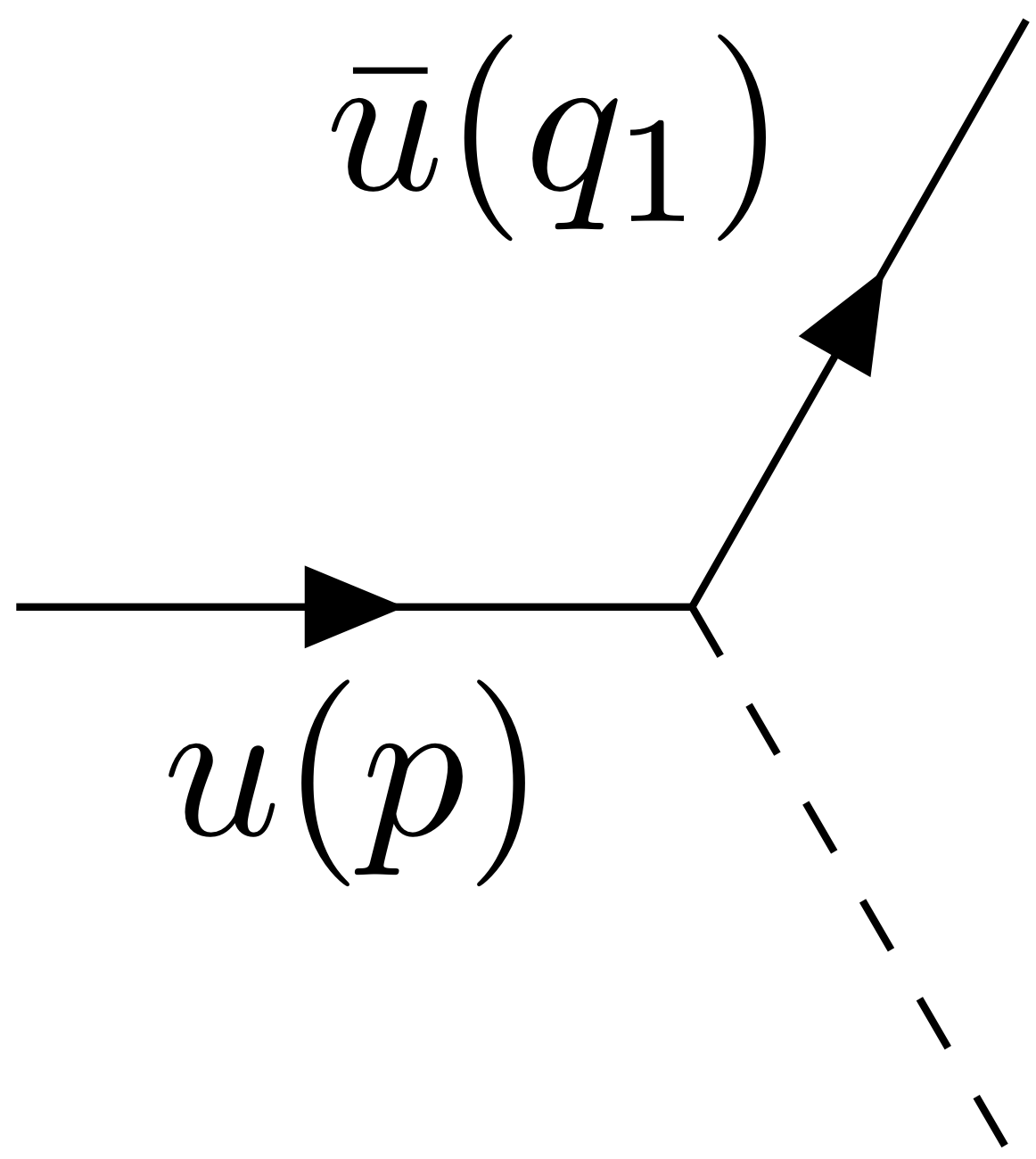}}
    \right] 
    = \bar{u}(q_1) \big[ -i(\kappa + \Tilde{\kappa} \gamma_5) \big] u(p),
\end{equation}
where we use the onshell incoming parton momentum ($p$), rather than the offshell momentum ($q_0$), not to spoil the kinematics of the incoming partons. $\kappa$ and $\Tilde{\kappa}$ are the CP-even and the CP-odd couplings, respectively.

\begin{table}[!h]
\centering
\setlength\extrarowheight{4pt}
\begin{tabular}{|| c || c | c ||}
    \hline
    $\mathcal{M}_{\lambda_0,\ \lambda_1}^{f\rightarrow f' \phi}$ & \multicolumn{2}{c||}{$\lambda_1$} \\ [0.5ex] 
    \hline
    $\lambda_0$ & $\uparrow$ & $\downarrow$ \\ [0.5ex]
    \hhline{||=||=|=||}
    $\uparrow$ & 
    $\frac{\kappa(zm_0+m_1)-\Tilde{\kappa}(zm_0-m_1)}{\sqrt{z}}$ & 
    $-\frac{(\kappa+\Tilde{\kappa})p_T}{\sqrt{z}}$  \\ [1ex] 
    \hline
    $\downarrow$ & 
    $\frac{(\kappa-\Tilde{\kappa})p_T}{\sqrt{z}}$ & 
    $\frac{\kappa(zm_0+m_1)+\Tilde{\kappa}(zm_0-m_1)}{\sqrt{z}}$  \\ [1ex] 
    \hline
\end{tabular}
\caption{Matrix elements of $f\rightarrow f'\phi$ splitting functions, where all phase terms, $e^{i(\lambda_0-\lambda_1)\phi}$, are factored out for simplicity.}
\label{t:ffs ME}
\end{table} 
With simple calculations, we can get the spin-unaveraged matrix elements $\mathcal{M}_{\lambda_0,\ \lambda_1}^{f\rightarrow f' \phi}$ as shown in table~\ref{t:ffs ME}. It shows an interesting symmetry from parity as follows:
\begin{itemize}
    \item Parity transformation~\cite{2_loop_sf, mhv_rules} - The splitting function of a total helicity flipped process should satisfy
    $$
        \mathcal{M}_{\lambda_0, \lambda_1}^{f\rightarrow f'\phi} = (-1)^{1+s_1+s_2} \left( \mathcal{M}_{-\lambda_0, -\lambda_1}^{f\rightarrow f'\phi} \right) ^* ,
    $$
    where $s_i$ is a spin of the $i^{\texttt{th}}$ fermion given by $s_i = \pm 1/2$.
\end{itemize}

The sum of the matrix element squares weighted by the spin density matrix can be written as
\begin{equation}
\begin{split}
    \underset{\textrm{pol}}{\resizebox{0.4cm}{!}{$\overline{\sum}$}}&|\mathcal{M}|^2 = \rho_+ |\mathcal{M}_{\uparrow \uparrow}|^2 + \rho_+ |\mathcal{M}_{\uparrow \downarrow}|^2 + \rho_- |\mathcal{M}_{\downarrow \uparrow}|^2 + \rho_- |\mathcal{M}_{\downarrow \downarrow}|^2 \\
    &= \rho_+ \frac{\big| \kappa(zm_0+m_1)-\Tilde{\kappa}(zm_0-m_1) \big|^2 + |\kappa+\Tilde{\kappa}|^2p_T^2}{z} \\
    &\quad + \rho_- \frac{\big| \kappa(zm_0+m_1)+\Tilde{\kappa}(zm_0-m_1) \big|^2 + |\kappa-\Tilde{\kappa}|^2p_T^2}{z} \\
    &= \left( \rho_+|\kappa+\Tilde{\kappa}|^2+\rho_-|\kappa-\Tilde{\kappa}|^2 \right) \big[z(1-z)^2\Tilde{q}^2-m_2^2\big] + (\rho_+ + \rho_-) \big[ |\kappa|^2(m_0+m_1)^2\\
    &\quad +|\Tilde{\kappa}|^2(m_0-m_1)^2 \big] + 2(\rho_+ - \rho_-) \Re(\kappa\Tilde{\kappa}^*) \big[(1-2z)m_0^2+m_1^2\big],
\end{split}
\end{equation}
where ${\resizebox{0.4cm}{!}{$\overline{\sum}$}}$ means that the sum is weighted by the spin density of the incoming parton, and $\Re(\cdot)$ stands for the real part of a complex number. Hence, $\rho_+$ and $\rho_-$ are the first and second diagonal elements of the spin density matrix. We use $p_T^2 = z^2(1-z)^2 \Tilde{q}^2 + z(1-z)m_0^2 - (1-z)m_1^2 - zm_2^2$ to derive the last line.

Finally, the splitting probability takes on the following form:
\begin{equation}
\begin{split}
    &P_{f\rightarrow f'\phi}(z,\ \Tilde{q}) = \frac{1}{2(q_0^2-m_0^2)} \underset{\textrm{pol}}{\resizebox{0.4cm}{!}{$\overline{\sum}$}}|\mathcal{M}|^2 \\
    &= \frac{g^2}{2} \Big[
    \left( \rho_+|\kappa+\Tilde{\kappa}|^2+\rho_-|\kappa-\Tilde{\kappa}|^2 \right) \big[(1-z)-m_{2,t}^2\big] + (\rho_+ + \rho_-) \big[ |\kappa|^2(m_{0,t}+m_{1,t})^2 \\
    &\quad +|\Tilde{\kappa}|^2(m_{0,t}-m_{1,t})^2 \big] + 2(\rho_+ - \rho_-) \Re(\kappa\Tilde{\kappa}^*) \big[(1-2z)m_{0,t}^2+m_{1,t}^2\big] \Big],
\end{split}
\label{eq:ffs splitting function}
\end{equation}
where $m_{i,t} = \frac{m_i}{\sqrt{z(1-z)}\Tilde{q}}$ is used for convenience.

\textsf{Herwig7} does not treat leptons as progenitors because of the complexities and potential divergences associated with lepton splittings. For instance, while we can mitigate divergence in coloured particles by assigning a constituent mass as a minimum cut, such a workaround does not exist for leptons. As a result, \textsf{Herwig7} does not handle processes like $\ell \rightarrow \ell'\phi$, meaning lepton-induced processes such as $\ell \rightarrow \ell Z'$ are not included. 

In the context of squark, gluino, and quark splitting functions, ignoring all mass terms yields results consistent with those described in Appendix C of ref.~\cite{SUSY_splitting}. For example, the $\Tilde{g}\rightarrow q\Tilde{q}$ splitting probability is given by $P_{\Tilde{q}_{L/R}\Tilde{g}}(x) = \frac{1}{2}T_fx$, and the $\Tilde{g}\rightarrow q\Tilde{q}_{L/R}$ splitting probability is $P_{q\Tilde{g}}(x) = 2\cdot \frac{1}{2} T_f(1-x)$. These are special cases of Eq.~\eqref{eq:ffs splitting function} with $\kappa = 1$, $\Tilde{\kappa} = 0$, $\rho_0+\rho_1=1$, matching the colour factors of the corresponding Standard Model splittings, and an additional factor of $2$ in the second function due to the sum of the $\Tilde{g}\rightarrow q\Tilde{q}_L$ and $\Tilde{g}\rightarrow q\Tilde{q}_R$ splittings.

For the SM coupling with $g = g_W m_0/2m_W$, where 
$$
g_W = e/\sin{\theta_W}, \quad 
m_0 = m_1, \quad 
\kappa = 1, \quad 
\Tilde{\kappa} = 0, 
\quad \rho_0+\rho_1=1,
$$
the splitting probability is expressed as:
\begin{equation}
    P_{f\rightarrow f\phi}^{SM}(z, \Tilde{q}) = \frac{g_W^2}{8} \Big( \frac{m_0}{m_W} \Big)^2 \bigg[ (1-z) + \frac{4m_0^2-m_2^2}{z(1-z)\Tilde{q}^2} \bigg].
\end{equation}
This formulation successfully replicates the EW splitting result as described in Eq. (3.21) of ref.~\cite{herwig_ewk_radiation}. A noteworthy aspect of the BSM $f\rightarrow f'\phi$ splitting is the possibility of the incoming and outgoing fermions having different flavours. This is reflected in the appearance of the $m_1$ term in eqn.~\ref{eq:ffs splitting function}.

\subsection{\texorpdfstring{$V\rightarrow V'\phi$}{V to V'phi} splitting function}
\label{sec:vvh}

\begin{equation}
    -i\mathcal{M}\left[
    \raisebox{-0.38in}{\includegraphics[clip, width=0.12\textwidth]{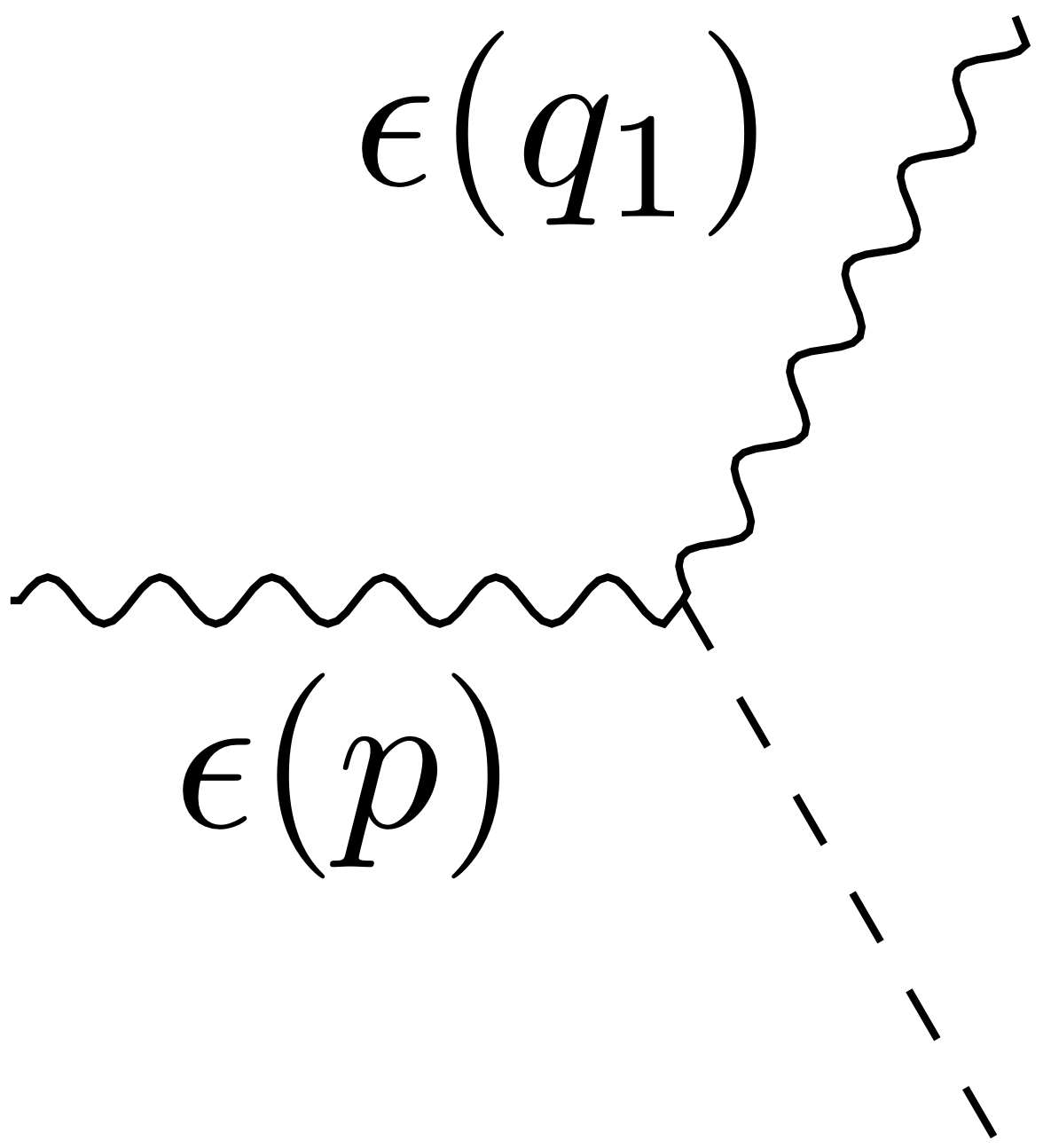}}
    \right] 
    = \epsilon^{*\mu}(q_1) (ig_{BSM}g_{\mu\nu}) \epsilon^{\nu}(p).
\label{eq:vvh Feynman}
\end{equation}

In the $V\rightarrow V \phi$ splittings, the SM vertex factor is defined as $im_V g_V g^{\mu\nu}$, with 
$$
g_W = e/\sin\theta_W, \qquad g_Z = e/\sin\theta_W \cos\theta_W.
$$
This vertex factor inherently includes a mass term. A similar approach is used for the BSM vertices, where the mass term is typically incorporated into the BSM coupling constant, denoted as $g_{BSM} (\sim m g')$, indicating an order of mass. Therefore, for consistency, it is essential to compute $\mathcal{M}/g_{BSM}$ up to the 0th order in the small quantities of the quasi-collinear approximation, $p_T$ and $m_i$, which results in a cross section to the 2nd order. By employing the matrix element defined in Eq.~\eqref{eq:vvh Feynman}, Table~\ref{t:vvh ME} is obtained, showcasing the matrix elements of $V\rightarrow V'\phi$ splittings for all vector boson polarization combinations. This table illustrates the symmetry arising from parity, expressed as $\mathcal{M}_{\lambda_0, \lambda_1}^{V\rightarrow V'\phi} = (-1)^{\lambda_0+\lambda_1} \left( \mathcal{M}_{-\lambda_0, -\lambda_1}^{V\rightarrow V'\phi} \right) ^*$~\cite{2_loop_sf, mhv_rules}.

\begin{table}[!h]
\centering
\begin{tabular}{|| c || c | c | c | c ||}
    \hline
    $\mathcal{M}_{\lambda_0,\ \lambda_1}^{V\rightarrow V' \phi}/ig$ & \multicolumn{4}{c||}{$\lambda_1$} \\[0.5ex]
    \hline
    $\lambda_0$ & + & - & 0 & $0^*$  \\ [0.5ex]     
    \hhline{||=||=|=|=|=||}
    
    + & 
    \scalebox{0.8}{$-1$} & 
    \scalebox{0.8}{$0$} & 
    $\frac{p_T}{\sqrt{2} m_1}$ & 
    - \\ [1ex]
    \hline
    
    -               & 
    \scalebox{0.8}{$0$} & 
    \scalebox{0.8}{$-1$} & 
    $-\frac{p_T}{\sqrt{2} m_1}$ & 
    - \\ [1ex]
    \hline
    
    0               & 
    $-\frac{p_T}{\sqrt{2} z m_0}$ & 
    $\frac{p_T}{\sqrt{2} z m_0}$ & 
    $\frac{-z^2m_0^2-m_1^2+p_T^2}{2zm_0m_1}$ & 
    - \\ [1ex]
    \hline
    
    $0^*$             & 
    - & 
    - & 
    - &
    \scalebox{0.8}{$0$} \\ [1ex]
    \hline
\end{tabular}
\caption{Matrix elements of $V\rightarrow V'\phi$ splitting functions, where a coupling constant, $g$, and phase terms, $e^{i(\lambda_0-\lambda_1)\phi}$, are factored out.}
\label{t:vvh ME}
\end{table}

While we can eliminate the single pole phenomena in $\mathcal{M}^{V\rightarrow V'\phi}_{\pm, 0}$ or $\mathcal{M}^{V\rightarrow V'\phi}_{0, \pm}$ using an additional mass term in the coupling constant, a double pole still appears in the $(\lambda_0, \lambda_1) = (0, 0)$ branching. To address this, we apply Dawson's approach~\cite{Dawson}, and the results are denoted as $0^*$. It is important to note that Dawson's method is applied only in cases where the splitting function exhibits a double pole.

Putting the above notes together, the matrix element for a generic $V\rightarrow V \phi$ splitting can be calculated as
\begin{equation}
\begin{split}
    \underset{\textrm{pol}}{\resizebox{0.4cm}{!}{$\overline{\sum}$}}|\mathcal{M}|^2 &= \rho_+ \big( |\mathcal{M}_{++}|^2 + |\mathcal{M}_{+0}|^2 \big) + \rho_- \big( |\mathcal{M}_{--}|^2 + |\mathcal{M}_{-0}|^2 \big) \\
    &\ \ \ \ \ \ \ \ \ \ + \rho_0 \big( |\mathcal{M}_{0+}|^2 + |\mathcal{M}_{0-}|^2 + |\mathcal{M}_{0^*0^*}|^2 \big) \\
    &= g_{BSM}^2 \bigg[ \rho_+ \bigg(1 + \frac{p_T^2}{2m_1^2} \bigg) + \rho_- \bigg(1 + \frac{p_T^2}{2m_1^2} \bigg) + \rho_0 \bigg(\frac{p_T^2}{z^2m_0^2} \bigg) \bigg] \\
    &= g_{BSM}^2 \bigg[ \frac{\rho_+ + \rho_-}{2m_1^2} \Big( z^2(1-z)^2\Tilde{q}^2 + z(1-z)m_0^2 + (1+z)m_1^2 -zm_2^2 \Big) \\ 
    &\ \ \ \ \ \ \ \ \ \ \ \ + \frac{\rho_0}{z^2m_0^2} \Big( z^2(1-z)^2\Tilde{q}^2 + z(1-z)m_0^2 - (1-z)m_1^2 - zm_2^2 \Big) \bigg],
\end{split}
\end{equation}
where $\rho_{+,0,-}$'s are the first, the second, and the third diagonal elements of the spin density matrix respectively. This gives the splitting probability for the $V\rightarrow V'\phi$ branching with the convention of $m_{i,t}^2 = m_i^2/(\Tilde{q}^2z(1-z))$ as
\begin{equation}
\begin{split}
    P_{V\rightarrow V'\phi}(z,\Tilde{q}) = \frac{g_{BSM}^2}{2} &\bigg[ \frac{\rho_+ + \rho_-}{2m_1^2} \Big( z(1-z) + z(1-z)m_{0,t}^2 + (1+z)m_{1,t}^2 - zm_{2,t}^2 \Big) \\
    &+ \frac{\rho_0}{z^2m_0^2} \Big( z(1-z) + z(1-z)m_{0,t}^2 - (1-z)m_{1,t}^2 - zm_{2,t}^2 \Big) \bigg].
\label{eq:vvh splitting function}
\end{split}
\end{equation}
Here we want to re-emphasize that the coupling constant for $VV\phi$ vertices should have an order of mass so that $m_{0, 1}^2$ terms in the denominators do not diverge even though $p_T/m \rightarrow \infty$.

The SM splitting probability can be obtained by setting $m_0=m_1$ and replacing $g_{BSM}$ with $g_{SM}m_0$. This yields the following expression:
\begin{equation}
\begin{split}
    P_{V\rightarrow V\phi}(z,\Tilde{q}) = g_{SM}^2 &\bigg[ 
    \frac{1-z}{4z} \Big( (\rho_+ + \rho_-)z^2 + 2\rho_0 \Big)
    -\frac{m_{2,t}^2}{4z} \Big( (\rho_+ + \rho_-)z^2 + 2\rho_0 \Big) \\
    &-\frac{m_{0,t}^2}{4z^2} \Big( \rho_0(2z^2 - 4z + 2) + (\rho_+ + \rho_-)(z^4 - 2z^3 - z^2) \Big)
    \bigg].
\end{split}
\end{equation}
As anticipated, all singularities are successfully eliminated~\cite{herwig_ewk_radiation}. The notable difference between the generalized spin-0 particle radiation and the SM Higgs radiation from vector bosons is again the former can induce a flavour change in the bosonic flow leading to the emergence of the $m_1$ term in the function.

Despite the inclusion of an additional mass term in the coupling constant, Eq.~\eqref{eq:vvh splitting function} numerically diverges for massless vector bosons. A distinctive feature of BSM theories regarding the $VV\phi$ interactions is the capability of massless particles to interact with spin-0 particles, such as a $\gamma Z h$ interaction in the 2HDM. Therefore, it is prudent to individually consider massless splittings to properly address the singular behaviours arising from $1/m$ terms. When the incoming parton is massless, the splitting function omits the component derived from the longitudinal polarization of the incoming parton. Consequently, the splitting probability for a massless incoming vector boson is:
\begin{equation}
P_{V_{\text{massless}}\rightarrow V'\phi}(z,\Tilde{q}) = \frac{g_{BSM}^2}{4m_1^2} \left[ z(1-z) + (1+z)m_{1,t}^2 - zm_{2,t}^2 \right],
\end{equation}
noting that $\rho_+ + \rho_-$ is invariably unity for a massless incoming parton. The second scenario involves a massless outgoing vector boson, namely $V\rightarrow V'_{\text{massless}}\phi$. In this case, terms corresponding to $\lambda_1 = 0$ should be excluded, leading to a modified splitting probability:
\begin{equation}
\begin{split}
P_{V\rightarrow V'_{\text{massless}}\phi}(z,\Tilde{q}) = \frac{g_{BSM}^2}{2m_0^2} \bigg[ (\rho_+ + \rho_-) m_{0,t}^2 + \frac{\rho_0}{z^2} \left( z(1-z) + z(1-z)m_{0,t}^2 - zm_{2,t}^2 \right) \bigg].
\end{split}
\end{equation}

Finally, let us consider a scenario where both the incoming and outgoing vector bosons are massless. In general, scalar particles do not interact when both incoming and outgoing particles are massless. However, some limiting cases could be conceivable at future colliders operating at much higher energy scales, where vector bosons are exceedingly light. This limiting scenario yields the following splitting probability:
\begin{equation}
P_{V_{\text{massless}}\rightarrow V_{\text{massless}}'\phi}(z,\Tilde{q}) = \frac{g_{BSM}^2}{2z(1-z)\Tilde{q}^2}.
\end{equation}
It is important to note that this resulting formula aligns exactly with the $\phi\rightarrow \phi'\phi''$ splitting probability expressed in Eq.~\eqref{eq:sss splitting function}. This indicates that in the context of our universal analysis, the vector bosons involved in the $V\rightarrow V'\phi$ splitting assume the properties of Goldstone modes in the massless limit, despite the exclusion of longitudinal polarization contributions.

\section{Vector boson splittings}
\label{sec:vector}

In this section, we detail the emission dynamics for vector bosons off scalar, fermionic, and bosonic flows. The $\phi \rightarrow \phi' V$ splitting represents a pure BSM phenomenon, and thus we introduce this splitting function here. Following this, the generalized $f\rightarrow f'V$ and $V\rightarrow V'V''$ splitting functions are presented. Unlike scalar particle splittings, SM vector boson splittings have already introduced a flavour-changing current through the CKM matrix. This implies that for the $f\rightarrow f'V$ and $V\rightarrow V'V''$ processes, there isn't a significant distinction between the EW vector boson splitting functions~\cite{herwig_ewk_radiation} and their BSM counterparts. The notable difference in BSM theories is again the occurrence of FCNC. We have therefore calculated the splitting functions to be as general as possible, compared them with the EW functions~\cite{herwig_ewk_radiation}, and closely examined their properties.

\subsection{\texorpdfstring{$\phi\rightarrow \phi'V$}{phi to phi'V} splitting}
\label{sec:hhv}

The $\phi\rightarrow \phi'V$ branching is a distinctive aspect of BSM theories like 2HDM or SUSY. For the $\phi\phi V$ coupling, the vertex factor is represented as $ig(p_1-p_2)_{\mu}$, where all momenta are directed inward. This leads to the following matrix element:
\begin{equation}
    -i\mathcal{M}\left[
    \raisebox{-0.38in}{\includegraphics[clip, width=0.17\textwidth]{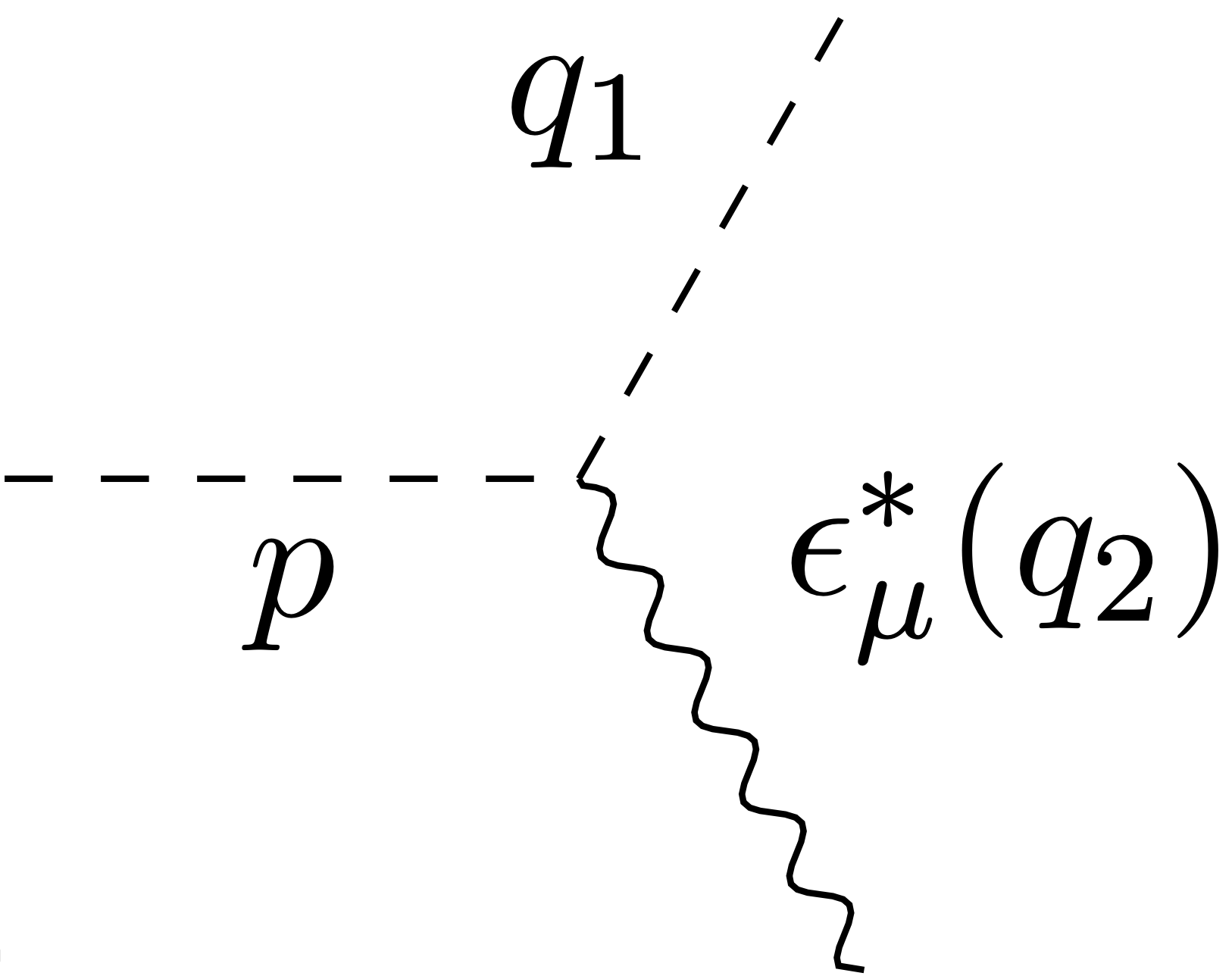}}
    \right] 
    = ig (p+q_1)^{\mu} \epsilon_{\mu}^*(q_2).
\end{equation}
By substituting Eqs.~\eqref{eq:qi}, \eqref{eq:polarization_qi}, and \eqref{eq:polarization_qi*}, the matrix elements are detailed in table~\ref{t:hhv ME}. When integrating these terms and employing the abbreviation for mass terms described in Eq.~\eqref{sec:ffh}, the splitting probability for a $\phi \rightarrow \phi'V$ branching is formulated as follows:
\begin{equation}
P_{\phi \rightarrow \phi'V}(z,\Tilde{q}) = g^2 \Bigg[
\frac{2z}{1-z} (1+m_{0,t}^2) - \frac{2}{1-z}m_{1,t}^2 + \frac{1}{2}m_{2,t}^2
\Bigg].
\end{equation}
Significantly, this splitting probability aligns with the massless SUSY $\Tilde{q} \rightarrow \Tilde{q}g$ splitting function, disregarding all mass terms, expressed as $P_{\Tilde{q}\rightarrow \Tilde{q}g}(z) = 2 C_F z/(1-z)$, as documented in~\cite{SUSY_splitting}.

\begin{table}[!h]
\centering
\begin{tabular}{|| c || c | c | c ||}
    \hline
    $\lambda_2$ & + & - & $0^*$ \\ [0.5ex] 
    \hline \hline
    $\mathcal{M}_{\lambda_2}^{\phi\rightarrow \phi' V}$ & 
    $-\sqrt{2}\frac{p_T}{1-z}$ & 
    $\sqrt{2}\frac{p_T}{1-z}$ & 
    $-\frac{1+z}{1-z}m_2$\\ [0.5ex]
    \hline
\end{tabular}
\caption{Matrix elements for the $\phi\phi V$ splitting functions (denoted as $\mathcal{M}_{\lambda_2}^{\phi \rightarrow \phi 'V}$) are presented, where the coupling constant $g$, and phase factors $e^{-i\lambda_2}$, have been factored out.}
\label{t:hhv ME}
\end{table} 

For $\phi\phi V$ couplings, the $V\rightarrow \phi \phi'$ branchings are also among notable phenomena in BSM theories. A splitting function for this process can be derived using the following symmetry properties:
\begin{itemize}
    \item Crossing symmetry~\cite{bible} - Swapping two final state particles entails a change $\big( z \Leftrightarrow (1-z), m_1 \Leftrightarrow m_2 \big)$, along with appropriate adjustments to the phase factors. Thus, 
    $$
    \mathcal{M}_{\lambda_0, \lambda_1, \lambda_2}^{A\rightarrow BC}(z,\Tilde{q}; m_0, m_1, m_2) \propto \mathcal{M}_{\lambda_0, \lambda_2, \lambda_1}^{A\rightarrow CB}(1-z,\Tilde{q}; m_0, m_2, m_1).
    $$

    \item Drell-Levy-Yan crossing relation~\cite{DLY-1,DLY-2,DLY-NLO} - Switching the incoming and outgoing partons results in
    $$
    P^{A\rightarrow BC}_{\lambda_A, \lambda_B, \lambda_C}(z) = (-1)^{\lambda_A+\lambda_B+\lambda_C}zP^{B\rightarrow AC}_{\lambda_B, \lambda_A, \lambda_C}(1/z).
    $$
\end{itemize}
Notably, for the $\phi\rightarrow \phi' V$ process, one needs to sum all polarization states of the final vector boson. However, for the $V\rightarrow \phi \phi'$ process, an average should be taken, as the vector boson now enters the vertex. The resulting splitting probability is:
\begin{equation}
\begin{split}
    P_{V\rightarrow \phi \phi'}(z,\Tilde{q}) = g^2 \Bigg[ 
    &(\rho_{+}+\rho_{-}) \left( z(1-z) (1+m_{0,t}^2) - (1-z)m_{1,t}^2 - z m_{2,t}^2 \right) \\
    &+ \rho_0 \frac{(1-2z)^2}{2}m_{0,t}^2
    \Bigg].
\end{split}
\end{equation}
It can be observed that this function aligns with the massless SUSY $g \rightarrow \Tilde{q} \Bar{\Tilde{q}}$ splitting function, $P_{g \rightarrow \Tilde{q} \Bar{\Tilde{q}}}(z) = T_f z (1-z)$~\cite{SUSY_splitting}, when $\rho_{+} + \rho_{-} = 1$ and $m_{0,1,2} \rightarrow 0$. The $V\rightarrow \phi\phi'$ splitting, although technically viable, is not introduced as a part of this generalized parton shower implementation since it can be only realized for $m=0$ orbital angular moment of the parent gauge vector boson. This constraint significantly limits the applicability of such a splitting in a generalized parton shower algorithm, which aims to be as universally applicable as possible across various processes and conditions. 

\subsection{\texorpdfstring{$f \to f'V$}{f to f'V} splitting}
\label{sec:ffv}

The invariant matrix element of the vector boson radiation from a fermion is given as
\begin{equation}
    -i\mathcal{M}\left[
    \raisebox{-0.38in}{\includegraphics[clip, width=0.17\textwidth]{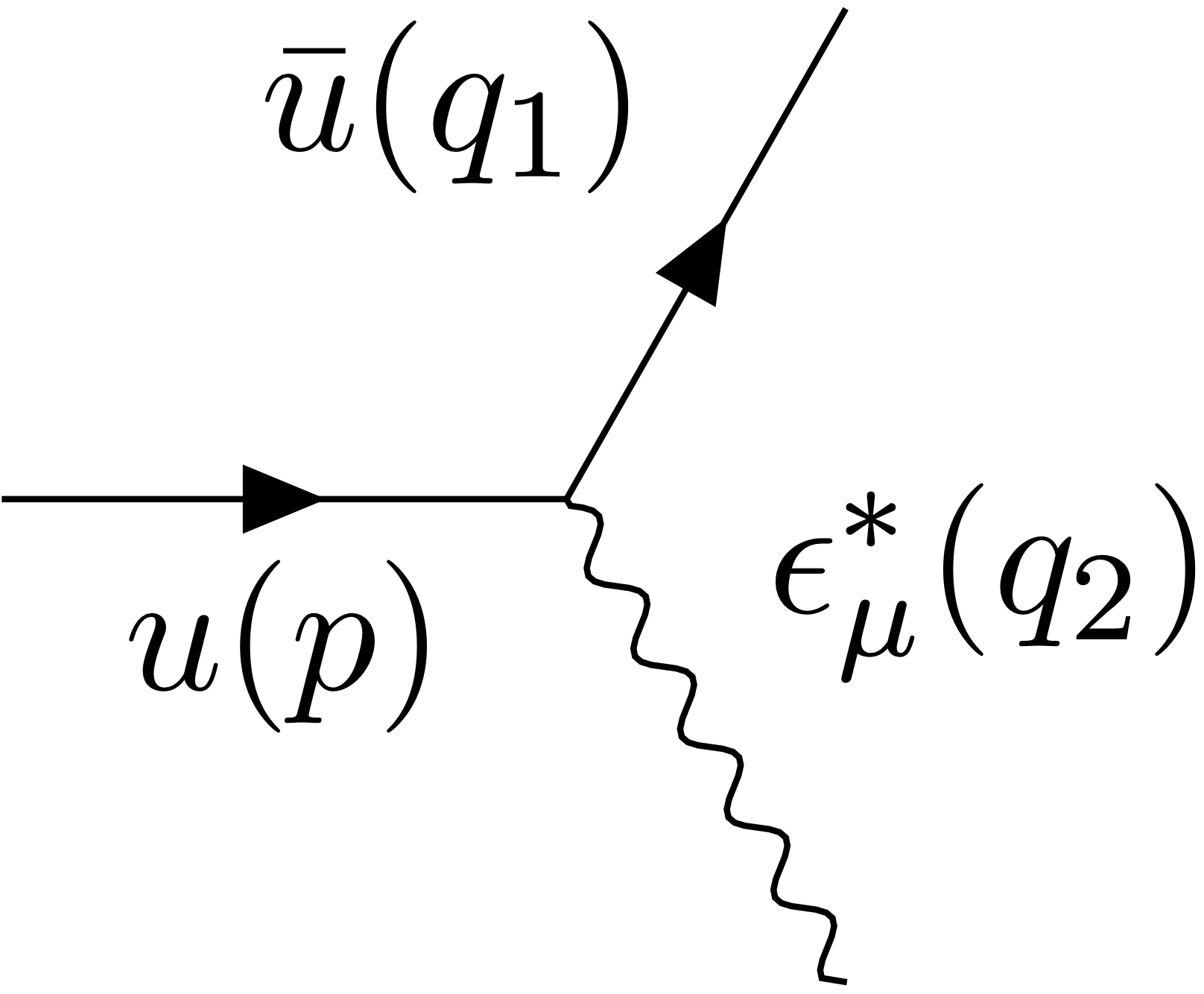}}
    \right] 
    = \bar{u}(q_1) \big[ -i(g_LP_L+g_RP_R) \gamma^\mu \big] \epsilon_\mu^*(q_2) u(p).
\end{equation}
The explicit forms of these spin-unaveraged matrix elements are given in Table~\ref{t:qqv ME}.

\begin{table}[!h]
\centering
\begin{tabular}{|| c | c || c | c | c | c ||}
    \hline
    \multicolumn{2}{||c||}{$\mathcal{M}_{\lambda_0,\lambda_1,\lambda_2}^{f\rightarrow f'V}$} & \multicolumn{4}{c||}{$\lambda_2$} \\ [0.5ex] 
    \hline
    $\lambda_0$ & $\lambda_1$ & + & - & 0 & $0^*$ \\ [0.5ex] 
    \hline\hline
    + & + & $\frac{\sqrt{2}g_Rp_T}{\sqrt{z}(1-z)}$ & $-\frac{\sqrt{2z}g_Rp_T}{1-z}$ & $\frac{g_L(1-z)^2m_0m_1+g_R(p_T^2-zm_2^2)}{\sqrt{z}(1-z)m_2}$ & $-\frac{2g_R\sqrt{z}m_2}{1-z}$\\ [0.5ex]
    + & - & $-\frac{\sqrt{2}(g_Lzm_0-g_Rm_1)}{\sqrt{z}}$ & $0$ & $-\frac{(g_Lm_0-g_Rm_1)p_T}{\sqrt{z}m_2}$ & 0 \\ [0.5ex]
    - & + & $0$ & $-\frac{\sqrt{2}(g_Rzm_0-g_Lm_1)}{\sqrt{z}}$ & $\frac{(g_Rm_0-g_Lm_1)p_T}{\sqrt{z}m_2}$ & 0 \\ [0.5ex]
    - & - & $\frac{\sqrt{2z}g_Lp_T}{1-z}$ & $-\frac{\sqrt{2}g_Lp_T}{\sqrt{z}(1-z)}$ & $\frac{g_R(1-z)^2m_0m_1+g_L(p_T^2-zm_2^2)}{\sqrt{z}(1-z)m_2}$ & $-\frac{2g_L\sqrt{z}m_2}{1-z}$\\ [0.5ex]
    \hline
\end{tabular}
\caption{Matrix elements of $f \to f'V$ splitting functions where phase terms $e^{i(\lambda_0-\lambda_1-\lambda_2)}$ are factored out.}
\label{t:qqv ME}
\end{table} 

To address the divergences arising from the longitudinal polarization, Dawson's approach, denoted as $0^*$, is again employed. The resulting splitting probability for $f\rightarrow f'V$ is:
\begin{equation}
\begin{split}
    P_{f\rightarrow f'V}(z,\Tilde{q}) = &(|g_R|^2\rho_+ + |g_L|^2\rho_-) \bigg( \frac{1+z^2}{1-z} (1+m_{0,t}^2) - \frac{1+z}{1-z}m_{1,t}^2 - m_{2,t}^2\bigg) \\
    &+ (|g_R|^2\rho_- + |g_L|^2\rho_+)zm_{0,t}^2 -2\Re(g_Lg_R^*)(\rho_+ + \rho_-)m_{0,t}m_{1,t}.
\label{eq:ffv splitting function}
\end{split}
\end{equation}
As anticipated, this formulation aligns precisely with the Standard Model (SM) $q\rightarrow q'V$ splitting function~\cite{herwig_ewk_radiation} when $g_L$ and $g_R$ are pure imaginary numbers, as is the case in the SM.

\subsection{\texorpdfstring{$V \to V'V''$}{V to V'V''} splitting}
\label{sec:vvv}

In this subsection, we discuss branchings of the $V$ to $V'V''$ process, where the invariant matrix element of this evolution can be calculated by
\begin{equation}
\begin{split}
    -i&\mathcal{M}\left[ 
    \raisebox{-0.38in}{\includegraphics[clip, width=0.17\textwidth]{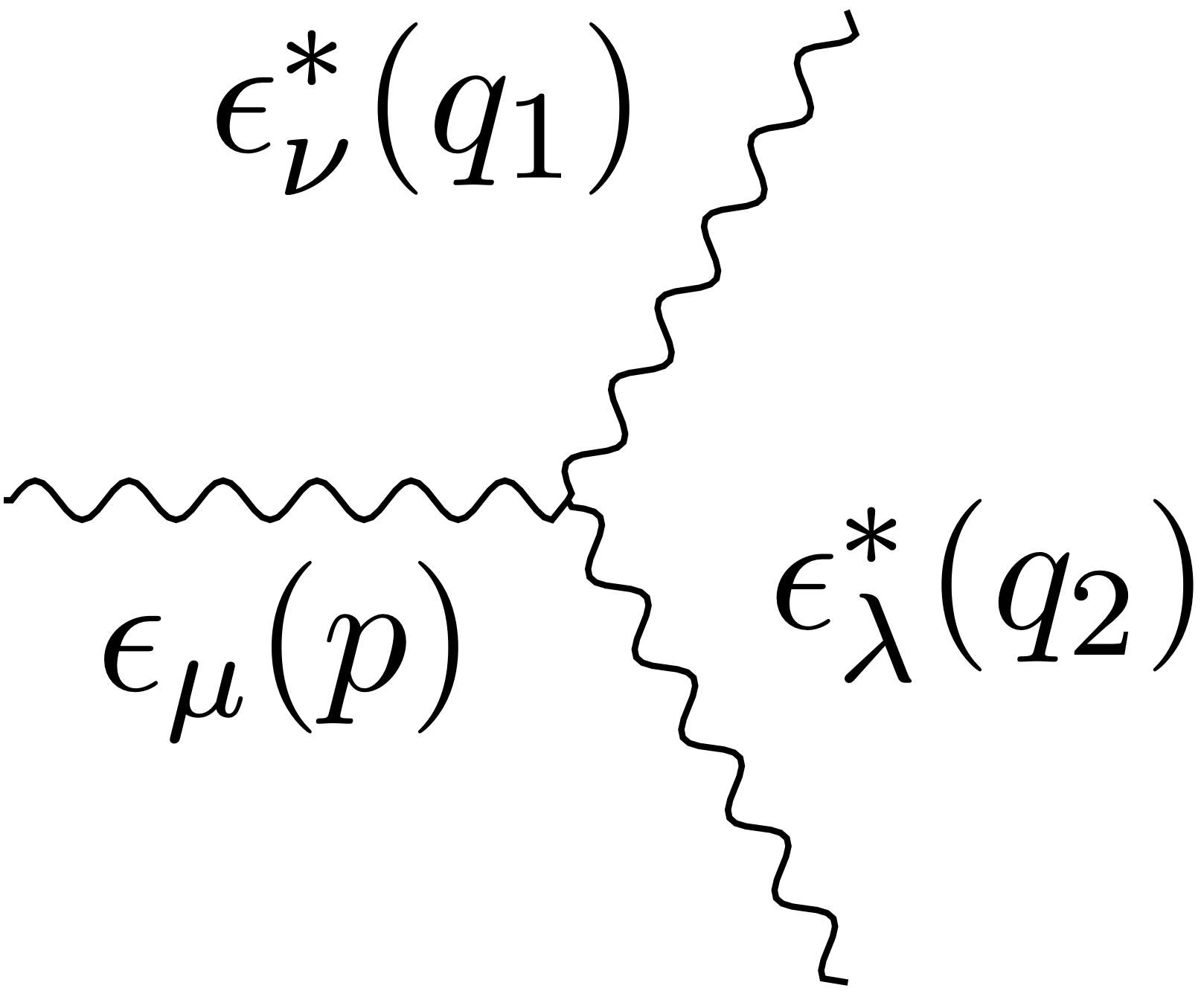}}
    \right] \\
    &= -g [g_{\mu\nu}(p+q_1)_\lambda + g_{\nu\lambda}(-q_1+q_2)_\mu - g_{\lambda\mu}(q_2+p)_\nu] \epsilon^\mu(p) \epsilon^{*\nu}(q_1) \epsilon^{*\lambda}(q_2).
\end{split}
\end{equation}
However, this formula cannot be used directly because the longitudinal polarization vectors revised by Dawson's approach are not orthogonal to the momentum vector, i.e.
\begin{equation}
    \epsilon_{0^*}(q_i) \cdot q_i = -q_i^2/m_i \not = 0.
\end{equation}
We therefore impose the orthogonality by requiring $\epsilon_{0^*}(q_i) \cdot q_i = 0$.
It is then handy to write down the matrix element in the following form:
\begin{equation}
\begin{split}
    -i\mathcal{M}_{V\rightarrow V'V''} = &-2g \left[ (q_1 \cdot \epsilon_2^*)(\epsilon_0 \cdot \epsilon_1^*) + (q_2 \cdot \epsilon_0)(\epsilon_1^* \cdot \epsilon_2^*) - (q_2 \cdot \epsilon_1^*)(\epsilon_0 \cdot \epsilon_2^*) \right] \\
    &+ g\ \beta_0 n\cdot \left[ \epsilon_2^* (\epsilon_0 \cdot \epsilon_1^*) + \epsilon_0 (\epsilon_1^* \cdot \epsilon_2^*) - \epsilon_1^* (\epsilon_0 \cdot \epsilon_2^*) \right].
\end{split}
\end{equation}
This formula can be employed for any polarization combination. It will give the explicit forms of matrix elements with regard to the polarizations of the particles that participated in the process given in table~\ref{t:vvv ME}, where we adopt Dawson's approach to deal with the divergence due to the longitudinal polarization states. Although the $V\rightarrow V'V''$ process is much more complex than the $\phi\rightarrow \phi'\phi''$ process, it exhibits all the aforementioned symmetric features such as the crossing symmetry, the Drell-Levy-Yan relation, and the parity transformation relation, since it is a maximally symmetric process as well. Finally, the splitting probability is given as
\begin{equation}
\begin{split}
    &P_{V\rightarrow V'V''}(z,\Tilde{q}) = 2g^2 \bigg[  
    \frac{ \left( 1-z(1-z) \right)^2 }{z(1-z)} (\rho_+ + \rho_-)
    + 2\rho_0 (1-z)^2m_{0,t}^2
    \\
    & + \frac{ \left( 1-z(1-z) \right)^2 m_{0,t}^2 - (1-z^2(1-z))m_{1,t}^2 - (1-z(1-z)^2)m_{2,t}^2}{z(1-z)} (\rho_+ + \rho_-)
    \bigg].
\end{split}
\end{equation}
It recovers the $g\rightarrow gg$ splitting function when all mass terms are neglected. 

\begin{table}[!h]
\centering
\rotatebox{90}{
\begin{tabular}{|| c | c || c | c | c | c ||}
    \hline
    \multicolumn{2}{||c||}{$\mathcal{M}_{\lambda_0,\lambda_1,\lambda_2}^{V\rightarrow V'V''}$} & \multicolumn{4}{c||}{$\lambda_2$} \\ [0.5ex]
    \hline
    $\lambda_0$ & $\lambda_1$ & + & - & $0$ & $0^*$ \\ [0.5ex] 
    \hhline{||=|=||=|=|=|=||}
    + & + & $-\frac{\sqrt{2}p_T}{z(1-z)}$ & $\frac{\sqrt{2}zp_T}{1-z}$ & $-\frac{z(1-z)(1+z)\Tilde{q}^2}{2m_2}+\frac{m_0^2}{m_2}-\frac{m_1^2}{m_2}-\frac{1+z}{1-z}m_2$ & $\frac{2zm_2}{1-z}$ \\ [1ex]
    + & - & \scalebox{0.8}{$=i\mathcal{M}_{++-}(z \Leftrightarrow (1-z))$} & $0$ & $0$ & $0$ \\ [1ex]
    - & + & $0$ & \scalebox{0.8}{$=-i\mathcal{M}_{++-}(z \Leftrightarrow (1-z))$} & $0$ & $0$ \\ [1ex]
    - & - & \scalebox{0.8}{$=-i\mathcal{M}_{++-}$} & \scalebox{0.8}{$=-i\mathcal{M}_{+++}$} & \scalebox{0.8}{$=i\mathcal{M}_{++0}$} & \scalebox{0.8}{$=i\mathcal{M}_{++0^*}$} \\ [1ex]
    \hline
    + & 0 & \scalebox{0.8}{$=-i\mathcal{M}_{++0}(z\Leftrightarrow (1-z), m_1 \Leftrightarrow m_2)$} & $0$ & $\left[ \frac{z(1-z)\Tilde{q}^2}{2\sqrt{2}m_1m_2}+\frac{m_0^2-m_1^2-m_2^2}{\sqrt{2}m_1m_2} \right] p_T$ & $-$ \\ [1ex]
    - & 0 & $0$ & \scalebox{0.8}{$=i\mathcal{M}_{+0+}$} & \scalebox{0.8}{$=-i\mathcal{M}_{+00}$} & $-$ \\ [1ex]
    + & $0^*$ & \scalebox{0.8}{$=-i\mathcal{M}_{++0^*}(z\Leftrightarrow (1-z), m_1 \Leftrightarrow m_2)$} & $0$ & $-$ & $0$ \\ [1ex]
    - & $0^*$ & $0$ & \scalebox{0.8}{$=i\mathcal{M}_{+0^*+}$} & $-$ & $0$ \\ [1ex]
    \hline
    0 & + & $0$ & $-\frac{z(1-z)(1-2z)\Tilde{q}^2}{2m_0}-\frac{(1-2z)m_0^2+m_1^2-m_2^2}{m_0}$ & $-\left[ \frac{z(1-z)\Tilde{q}^2}{2\sqrt{2}m_0m_2} + \frac{m_0^2-m_1^2-m_2^2}{\sqrt{2}zm_0m_2} \right] p_T$ & $-$ \\ [1ex]
    0 & - & \scalebox{0.8}{$=i\mathcal{M}_{0+-}$} & $0$ & \scalebox{0.8}{$=-i\mathcal{M}_{0+0}$} & $-$ \\ [1ex]
    $0^*$ & + & $0$ & \scalebox{0.8}{$-2(1-z)m_0$} & $-$ & $0$ \\ [1ex]
    $0^*$ & - & \scalebox{0.8}{$=i\mathcal{M}_{0^*+-}$} & $0$ & $-$ & $0$ \\ [1ex]
    \hline
    0 & 0 & \scalebox{0.8}{$=i\mathcal{M}_{0+0}(z \Leftrightarrow (1-z), m_1 \Leftrightarrow m_2)$} & \scalebox{0.8}{$=-i\mathcal{M}_{00+}$} & $\frac{\epsilon_{ijk}(z_i-z_j)(m_k^2p_T^2-z_k^2m_i^2m_j^2)}{4z_0z_1z_2m_0m_1m_2}$ & $-$ \\ [1ex]
    $0^*$ & $0^*$ & $0$ & $0$ & - & $0$ \\ [1ex]
    \hline
\end{tabular}
}
\caption{Matrix elements of $V \to V'V''$ splitting functions, where a coupling constant, $g$, and phase terms, $e^{i(\lambda_0-\lambda_1-\lambda_2)}$, are factored out.}
\label{t:vvv ME}
\end{table} 

\section{Results and discussions}
\label{sec:result}

In this section, we showcase the validation and outcomes from incorporating generalized splitting functions, encompassing BSM boson radiation schemes, into \textsf{Herwig~7}. Our study centres on an extensive performance assessment of BSM parton showers, covering an array of BSM scenarios, as well as the SM Triple Higgs Coupling (THC), a phenomenon within the SM framework that had not been implemented in the previous iterations of \textsf{Herwig~7}.

To rigorously validate the generalized parton shower scheme, several SM and BSM scenarios were meticulously chosen. We selected $n+1$ particle processes, each sensitive to distinct types of splittings, and ensured their phase spaces were minimally affected by interference with non-parton shower-like diagrams. For each process, we conducted two different calculations: one with and one without a final state emission resulting in the same Feynman diagram. Specifically, we first generated a hard matrix element (ME, denoted as $\mathcal{M}$) using \textsf{MadGraph5}, then simulated shower processes with \textsf{Herwig~7}, limiting the parton shower ($\mathcal{PS}$) to allow only one corresponding FS BSM emission, denoted as $\mathcal{M}_n + \mathcal{PS}_{n\rightarrow n+1}$. This approach enabled us to gather single-step resummation data (RS). These events were then compared to the corresponding $n+1$ particle processes generated by \textsf{MadGraph5}, i.e., $\mathcal{M}_{n+1}$, which represents the fixed-order (FO) contribution. The UFO model files~\cite{UFO_model_file}, based on the FeynRules~\cite{feynrules2_manual} framework, were input into both \textsf{MadGraph5} and \textsf{Herwig~7} to accurately calculate FO and RS data for BSM scenarios.

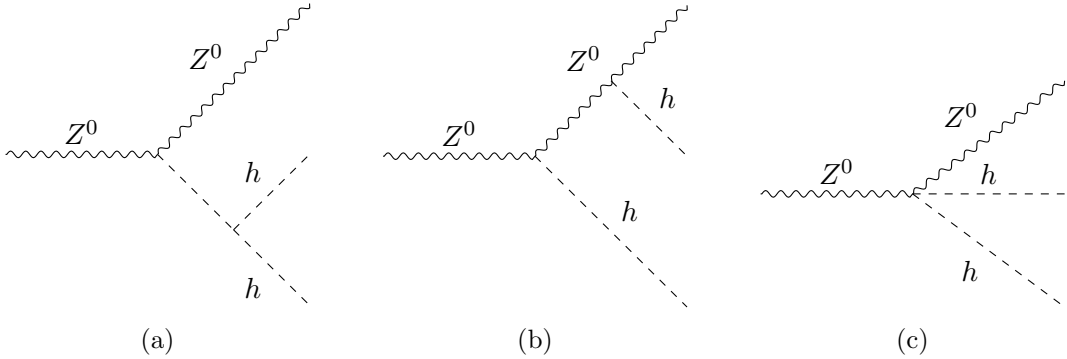
\begin{figure}
\centering
\begin{subfigure}{0.32\textwidth}
\centering
\begin{tikzpicture}[baseline=(current  bounding  box.base)]
\begin{feynman}
    \vertex (a1);
    \vertex [right=20mm of a1] (b1);
    \vertex [right=10mm of b1] (c1);
    \vertex [right=10mm of c1] (d1);
    \vertex [below=10mm of c1] (c2);
    \vertex [above=10mm of c1] (c3);
    \vertex [below=10mm of d1] (d2);
    \vertex [above=10mm of d1] (d3);
    \vertex [below=10mm of d2] (d4);
    \vertex [above=10mm of d3] (d5);

    \diagram* {
        (a1) -- [boson, edge label=\(Z^0\)] (b1) -- [boson, edge label=\(Z^0\)] (d5),
        (b1) -- [scalar] (c2) -- [scalar, edge label=\(h\)] (d1),
        (c2) -- [scalar, edge label'=\(h\)] (d4)
    };
\end{feynman}
\end{tikzpicture}
\caption{}\label{f:hhh emission}
\end{subfigure}
\begin{subfigure}{0.32\textwidth}
\centering
\begin{tikzpicture}[baseline=(current  bounding  box.base)]
\begin{feynman}
    \vertex (a1);
    \vertex [right=20mm of a1] (b1);
    \vertex [right=10mm of b1] (c1);
    \vertex [right=10mm of c1] (d1);
    \vertex [below=10mm of c1] (c2);
    \vertex [above=10mm of c1] (c3);
    \vertex [below=10mm of d1] (d2);
    \vertex [above=10mm of d1] (d3);
    \vertex [below=10mm of d2] (d4);
    \vertex [above=10mm of d3] (d5);

    \diagram* {
        (a1) -- [boson, edge label=\(Z^0\)] (b1) -- [boson, edge label=\(Z^0\)] (d5),
        (b1) -- [scalar, edge label=\(h\)] (d4),
        (c3) -- [scalar, edge label=\(h\)] (d1)
    };
\end{feynman}
\end{tikzpicture}
\caption{}\label{f:zzh double emission}
\end{subfigure}
\begin{subfigure}{0.32\textwidth}
\centering
\begin{tikzpicture}[baseline=(current  bounding  box.base)]
\begin{feynman}
    \vertex (a1);
    \vertex [right=20mm of a1] (b1);
    \vertex [right=20mm of b1] (c1);
    \vertex [below=15mm of c1] (c2);
    \vertex [above=15mm of c1] (c3);

    \diagram* {
        (a1) -- [boson, edge label=\(Z^0\)] (b1) -- [boson, edge label=\(Z^0\)] (c3),
        (b1) -- [scalar, edge label=\(h\)] (c1),
        (b1) -- [scalar, edge label'=\(h\)] (c2)
    };
\end{feynman}
\end{tikzpicture}
\caption{}\label{f:zzhh quartic coupling}
\end{subfigure}
\vspace*{-1em}
\caption{Feynman diagrams for the $Z \to Zhh$ process.}
\label{f:zhh diagram}
\end{figure}

For the $\phi \to \phi'\phi''$ splitting outlined in section~\ref{sec:sss}, the SM THC provides a straightforward example for examination. We utilize the $pp\rightarrow Z^0 h$ process (simulated with \textsf{MadGraph5}) as a basis for the RS contributions of the $h\rightarrow hh$ splitting (modelled in \textsf{Herwig~7}). Corresponding FO calculations are performed using the $pp \rightarrow Z^0 h \rightarrow Z^0hh$ process. It is noteworthy that we specifically configure this process in \textsf{MadGraph5} to generate a diagram featuring the triple Higgs coupling%
\footnote{One might worry that this choice is not gauge-invariant, but we have verified with \textsf{MadGraph5} that for this process, within the phase space dominated by the parton shower kinematics, interference with the non-emission diagrams is numerically small, so any associated ambiguities must also be small.}
(illustrated in figure~\ref{f:hhh emission}), while deliberately excluding irrelevant diagrams such as the double-splitting of Higgs bosons from a $Z$ boson (shown in figure~\ref{f:zzh double emission}) or a quartic coupling (depicted in figure~\ref{f:zzhh quartic coupling}).

All results plots are presented at the end of the paper, where figure~\ref{f:zhh} shows the results for the SM Higgs with mass 125 GeV at a centre-of-mass energy of $\sqrt{s} = 100$ TeV, chosen for its capability to facilitate a sufficient number of massive particle emissions. Panels~\subref{f:zhh_pt}, \subref{f:zhh_z}, \subref{f:zhh_mij}, and~\subref{f:zhh_drij} illustrate the differential rates of Higgs boson emissions as a function of the $h$ momentum transverse to the $hh$ axis ($p_T$), light-cone momentum fraction ($z$), the mass of the di-Higgs system ($m(h,h)$), and the angular separation between two Higgs bosons ($\Delta R(h,h)$), respectively. The red histogram represents the kinematics of the Higgs bosons emitted as a result of the newly implemented parton shower algorithms, labelled as RS, while the blue histogram corresponds to their FO counterparts.

The $h\rightarrow hh$ splitting RS shows a good correspondence with its FO counterpart. However, notable discrepancies are observed in regions characterized by high $p_T$, large $m(h,h)$, or substantial $\Delta R(h,h)$, where hard emissions predominate. In these areas, the collinear factorization theorem is less effective, but this issue can be addressed by integrating MEs that include additional hard jets. Furthermore, although the RS tends to slightly underestimate the extremities of the light-cone momentum fraction distribution, it aligns quite well at the central part, which contributes significantly to the overall results.


We further broadened our analysis to encompass BSM Higgs branchings within the general 2HDM. Figure~\ref{f:hmhp} displays the results for the $H \rightarrow h^+h^-$ branching, derived from the $p p \rightarrow H j$ process at a center-of-mass energy of $\sqrt{s} = 13.6$ TeV, reflecting the current  LHC energy configuration. In this context, $H$ represents an additional neutral CP-even Higgs boson, and $h^\pm$ denotes charged Higgs bosons characteristic of 2HDM, all set to a mass of 10 GeV. Heavy flavour quarks are excluded from jet constituents, thus eliminating the need to consider additional Higgs radiation from a jet. The panels exhibit the differential rates of emissions as functions of $p_T$, $z$, $\Delta R(j, h^+)$, and $\Delta R(h^+,h^-)$, respectively. The overall shapes are quite similar, largely due to the ratio of the Higgs bosons' mass to the energy scale ($m/\sqrt{s} \sim 0.001$) being almost the same as in the previous SM $h\rightarrow hh$ case. The reason the RS overestimates the FO result more than in the SM scenario is due to the smaller masses of the Higgs bosons in this case, which amplifies soft and collinear enhancements. Nonetheless, figure~\ref{f:hmhp_drri} confirms that the $\phi\rightarrow\phi'\phi''$ splitting function performs exceedingly well within the parton shower regime, supporting the notion that the $\phi\rightarrow \phi'\phi''$ RS is capable of reasonably describing the FO data.

\begin{figure}
\centering
\begin{subfigure}{0.49\textwidth}
\centering
\begin{tikzpicture}[baseline=(current  bounding  box.base)]
\begin{feynman}
    \vertex (a1) {$e^-$};
    \vertex [right=20mm of a1] (b1);
    \vertex [right=15mm of b1] (c1);
    \vertex [right=20mm of c1] (d1) {$\Bar{b}$};
    
    \vertex [below=15mm of a1] (a2);
    \vertex [below=15mm of a2] (a3) {$e^+$};
    \vertex [below=15mm of b1] (b2);
    \vertex [below=15mm of b2] (b3);
    \vertex [below=15mm of c1] (c2);
    \vertex [below=15mm of c2] (c3);   
    \vertex [below=15mm of d1] (d2) {$H$};
    \vertex [below=15mm of d2] (d3) {$b$};

    \vertex [right=10mm of c1] (e1);
    \vertex [below=7.5mm of e1] (e2);

    \diagram* {
        (a1) -- [fermion] (b2) -- [fermion] (a3),
        (b2) -- [boson, edge label'=\(Z^0/\gamma\)] (c2),
        (d1) -- [fermion] (c2) -- [fermion] (d3),
        (e2) -- [scalar] (d2)
    };
\end{feynman}
\end{tikzpicture}
\caption{}\label{f:bbh2 radiation}
\end{subfigure}
\begin{subfigure}{0.49\textwidth}
\centering
\begin{tikzpicture}[baseline=(current  bounding  box.base)]
\begin{feynman}
    \vertex (a1) {$e^-$};
    \vertex [right=20mm of a1] (b1);
    \vertex [right=15mm of b1] (c1);
    \vertex [right=20mm of c1] (d1) {$\Bar{b}$};
    
    \vertex [below=15mm of a1] (a2);
    \vertex [below=15mm of a2] (a3) {$e^+$};
    \vertex [below=15mm of b1] (b2);
    \vertex [below=15mm of b2] (b3);
    \vertex [below=15mm of c1] (c2);
    \vertex [below=15mm of c2] (c3);   
    \vertex [below=15mm of d1] (d2) {$b$};
    \vertex [below=15mm of d2] (d3) {$H$};       

    \vertex [right=10mm of c1] (e1);
    \vertex [below=7.5mm of e1] (e2);

    \diagram* {
        (a1) -- [fermion] (b2) -- [fermion] (a3),
        (b2) -- [boson, edge label'=\(Z^0/\gamma\)] (c2) -- [boson, edge label=\(Z^0/\gamma\)] (e2),
        (d1) -- [fermion] (e2) -- [fermion] (d2),
        (c2) -- [scalar] (d3)
    };
\end{feynman}
\end{tikzpicture}
\caption{}\label{f:bbh2 hard}
\end{subfigure}
\vspace*{-1em}
\caption{Representative Feynman diagrams for the $e^+e^- \rightarrow b\Bar{b}H$ process.}
\label{f:bbh2 diagram}
\end{figure}
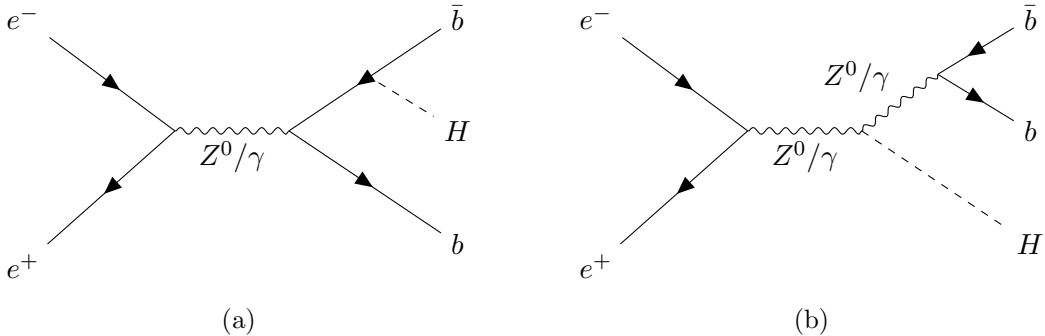

The performance testing of the $f\rightarrow f' \phi$ splitting, as detailed in section~\ref{sec:ffh}, utilizes the $e^+e^-\rightarrow b\Bar{b}$ process at a center-of-mass energy of $\sqrt{s} = 1$ TeV. Utilizing this hard process, single-step radiation of a $b\rightarrow bH$ emission and a $b\rightarrow sH$ emission are simulated separately with \textsf{Herwig~7}, where CP-even and -odd couplings are encompassed in both cases. The latter case specifically examines an FCNC process. Generating $e^+e^- \rightarrow qq'H$ with \textsf{MadGraph5} includes contributions from not only Final State Shower (FSS)-like diagrams, exemplified in figure~\ref{f:bbh2 radiation}, which are the target of our test, but also non-FSS-like diagrams such as those depicted in figure~\ref{f:bbh2 hard}. Consequently, figures~\ref{f:bbh2 mh2-10} and~\ref{f:bbh2 mh2-130} display the results for the $b\rightarrow bH$ emission, while figures~\ref{f:bsh2 mh2-10} and~\ref{f:bsh2 mh2-130} illustrate the outcomes of the $b\rightarrow sH$ emission. In each case, Higgs bosons with masses of 10 GeV and 130 GeV are simulated. These figures showcase the distributions of $\Tilde{q}$, $z$, $m(H, \text{branching partner})$, and $\Delta R(H, \text{branching partner})$, presented in sequence. The branching partner of the $H$ boson is an $s$-quark in the $b\rightarrow sH$ emission. However, selecting a branching partner in the $b\rightarrow bH$ case is more challenging. For the latter scenario, we compute the transverse momentum of the $H$ boson relative to both $b$- and $\Bar{b}$-quarks in the parton shower frame in which the $bH$ or $\Bar{b}H$ pair is on the axis, and select the quark that exhibits the lower transverse momentum.

The RS results match the shapes of the FO histograms. A notable decline at around $m(H, \text{branching partner}) \sim 500$ GeV in figure~\ref{f:bbh2 mh2-10 m} can be attributed to the maximal momentum constraint on the b-quark emanating from the 1 TeV $e^+e^-$ collision. This phenomenon is similarly observed in figures~\ref{f:bbh2 mh2-130},~\ref{f:bsh2 mh2-10}, and~\ref{f:bsh2 mh2-130}. Particularly noticeable are the sharp cutoffs in the $\Tilde{q}$ distributions at $\Tilde{q} = 1$ TeV, with minor remnants above this threshold resulting from energy-momentum redistribution during the final stages of parton showering, in line with the recoil scheme. This effect leads to more pronounced downturns at both extremes of the $z$ distributions, becoming more apparent in scenarios involving massive Higgs bosons. This is because the mass terms in equation~\ref{eq:ffs splitting function} invariably include a $1/z(1-z)\Tilde{q}$ factor. Moreover, by juxtaposing figures~\ref{f:bbh2 mh2-10 qt} and~\ref{f:bbh2 mh2-130 qt}, it is discerned that a smaller Higgs boson mass correlates with the initiation of parton radiation from lower $\Tilde{q}$, thereby allowing a greater expanse of phase space to be permitted. This phenomenon, notwithstanding the coupling constants being equivalent in both scenarios, amplifies the total cross section more considerably when the mass of the Higgs boson is smaller. This effect is notably accentuated at lower $\Delta R(b, H)$.
Finally, we note that the normalisation of the RS curves for SM $b(\Bar{b})\rightarrow b(\Bar{b})H$ is considerably higher than for FO. This comes about because the RS approach is able to use dynamical scale choices at vertices and, in the case of the $b\Bar{b}H$ vertex uses the running $b$ quark mass, which can be considerably larger at small scales than at the global event scale used in a FO calculation.


The $V\rightarrow V'\phi$ branching, as outlined in Section~\ref{sec:vvh}, is evaluated using a $pp\rightarrow W^\pm j$ underlying Born process, incorporating a $W^\pm \rightarrow W^\pm H$ emission, at a center-of-mass energy of 13.6 TeV. For this analysis, the mass of the $H$ boson is set at 1 GeV. To minimize interference from other diagrams yielding the same final states, the FO calculation is configured to specifically ensure that the $H$ boson is emitted from the $W$ boson. The results of this evaluation are depicted in figure~\ref{f:wh2j}, where the four panels display the same types of distributions as those shown in figure~\ref{f:bbh2 mh2-10}. It is evident from these results that all RS distributions exhibit trends that are similar to their FO counterparts, especially in the regions where the RS approach should work well: small $\Tilde{q}$ and $m$, and $\Delta R\lesssim\pi/2$.

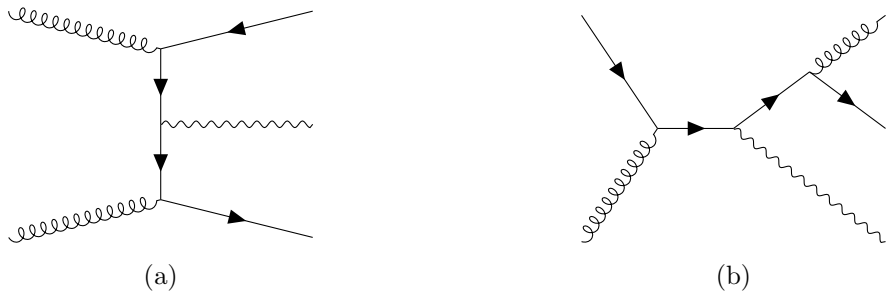
\begin{figure}
\centering
\begin{subfigure}{0.49\textwidth}
\centering
\begin{tikzpicture}[baseline=(current  bounding  box.base)]
\begin{feynman}
    \vertex (a0);
    \vertex [below=5mm of a0] (a1);
    \vertex [right=20mm of a1] (b1);
    \vertex [right=40mm of a0] (c1);
    \vertex [below=20mm of a1] (a2);
    \vertex [below=5mm of a2] (a3);
    \vertex [below=10mm of b1] (b2);
    \vertex [below=10mm of b2] (b3);
    \vertex [right=20mm of b2] (c2);
    \vertex [right=40mm of a3] (c3);
    \diagram* {
        (a0) -- [gluon] (b1),
        (a3) -- [gluon] (b3),
        (c1) -- [fermion] (b1) -- [fermion] (b2) -- [fermion] (b3) -- [fermion] (c3),
        (b2) -- [boson] (c2)
    };
\end{feynman}
\end{tikzpicture}
\caption{}\label{f:pp2jj qq-fusion}
\end{subfigure}
\begin{subfigure}{0.49\textwidth}
\centering
\begin{tikzpicture}[baseline=(current  bounding  box.base)]
\begin{feynman}
    \vertex (a0);
    \vertex [right=10mm of a0] (b0);
    \vertex [right=10mm of b0] (c0);
    \vertex [right=10mm of c0] (d0);
    \vertex [right=10mm of d0] (e0);
    \vertex [below=15mm of a0] (a1);
    \vertex [below=10mm of b0] (b1);
    \vertex [below=10mm of c0] (c1);
    \vertex [below=10mm of d0] (d1);
    \vertex [below=15mm of e0] (e1);
    \vertex [above=15mm of a0] (a-1);
    \vertex [above=10mm of b0] (b-1);
    \vertex [above=10mm of c0] (c-1);
    \vertex [above=7.5mm of d0] (d-1);
    \vertex [above=15mm of e0] (e-1);
    \diagram* {
        (a-1) -- [fermion] (b0) -- [fermion] (c0) -- [fermion] (d-1) -- [fermion] (e0),
        (a1) -- [gluon] (b0),
        (c0) -- [boson] (e1),
        (d-1) -- [gluon] (e-1)
    };
\end{feynman}
\end{tikzpicture}
\caption{}\label{f:pp2jj qv2qvg}
\end{subfigure}
\vspace*{-1em}
\caption{Non-FSS-like Feynman diagrams for the $pp\rightarrow Vjj$ process.}
\label{f:pp2jj diagram}
\end{figure}

We then proceeded to examine the $f\rightarrow f'V$ splitting function as described in section~\ref{sec:ffv}, under the conditions of the current LHC setup at $\sqrt{s} = 13.6$ TeV. To validate our results across different models, we utilized two distinct frameworks: firstly, the $q\rightarrow qZ'$ branching within the minimal B-L model, a $U(1)$ extension of the Standard Model featuring a gauged baryon-minus-lepton number~\cite{bl4-ufo-1, bl4-ufo-2, bl4-ufo-3}; and secondly, the $q\rightarrow q'W'$ branching in the $W'$ effective model~\cite{effW-ufo-1, effW-ufo-2}. To mitigate interference with non-FSS-like diagrams, as illustrated in figure~\ref{f:pp2jj diagram}, we exclusively selected incoming and outgoing quark flavours. The $Z'$ branching involved $u\Bar{u} \rightarrow d\Bar{d}$ processes, while the $W'$ model employed $u\Bar{u} \rightarrow c\Bar{c}$ as the underlying Born processes. Within \textsf{MadGraph5}, initial-state radiation contributions were eliminated using diagram filters. The masses for the $Z'$ and $W'$ bosons were set to 10 GeV and 50 GeV, respectively.

The results presented in Figures~\ref{f:ddzp} and~\ref{f:cswp} illustrate a generally accurate representation of the FO calculations. A notable peculiarity observed in these figures is a wiggle in the high $z$ region. This feature arises from the $1/(1-z)$ dependency of the mass terms in equation~\ref{eq:ffv splitting function}. As a consequence, figure~\ref{f:cswp z} exhibits more pronounced fluctuations compared to figure~\ref{f:ddzp z}, owing to the larger mass of the $W'$ boson. This effect is also evident in lab frame distributions like figure~\ref{f:cswp mij}. However, it does not significantly alter the overall trends in the results, indicating the robustness of the analysis in capturing the key dynamics of the processes under study.


The observed fluctuations in the plots for processes $f \rightarrow f'V$ and $V \rightarrow V'\phi$ are attributed to the dependency on the mass terms within the splitting functions, especially pronounced for bosons with significant mass such as the $W'$. These phenomena are indicative of the intricate dynamics at play when incorporating BSM physics into parton shower simulations, particularly under scenarios involving high transverse momentum or wide-angle emissions. The slight mismatches and the presence of wiggles, especially in regions of high $z$ and $p_T$, suggest the limitations of the collinear approximation, pointing towards the integration of matrix elements that include additional hard jets for enhanced precision in these regions. Moreover, the mass terms in the splitting functions significantly influence the kinematic distributions, such as $m(H, \text{branching partner})$ and $\Delta R$, by affecting the phase space and the probabilities of radiation at different scales. The dependence introduced by the mass terms, particularly through the $1/z(1-z)$ factor, is crucial for ensuring suppression in the forward region, underscoring the validity of the quasi-collinear approach. The comprehensive analysis across different models, including the minimal B-L model and the $W'$ effective model, and the comparison of RS to FO calculations, underline the effectiveness of the generalized parton shower scheme in capturing the essence of BSM radiation dynamics.

Figure~\ref{f:qqv} provides additional validation for the $q\rightarrow q'V$ splitting by comparing the $z$ distribution under the same conditions as presented in figures~\ref{f:ddzp z} and~\ref{f:cswp z}. The primary distinction between them lies in the masses of the $W'$ and $Z'$ bosons, with the current figures depicting scenarios where $m(W') = 10$ GeV (fig.~\ref{f:wp10}), $m(W') = 100$ GeV (fig.~\ref{f:wp100}), and $m(Z') = 200$ GeV (fig.~\ref{f:zp200}), respectively. Notably, consistent wiggles are observed near $z\rightarrow 1$ when the masses of the vector bosons are large, whereas this feature is absent in scenarios with lower boson masses, irrespective of the vector boson types.

\section{Conclusion}
\label{sec:conclusion}

In this study, we have presented a novel extension of the AO parton shower scheme to incorporate BSM splittings by examining generalized splitting functions. We began by systematically deriving explicit expressions of the quasi-collinear matrix elements in helicity-dependent forms. The splitting functions correspond well with preceding results~\cite{herwig_ewk_radiation,vincia_ewk_radiation,ewk_radiation-theory_GET,SUSY_splitting,SUSY_splitting-2} in the EW or SUSY limits.

Existing AO shower functions are augmented with these BSM splitting functions, which have been integrated into the shower process of \textsf{Herwig~7}. To ensure the accuracy and reliability of our implementation, we compared the results of the implemented BSM parton shower against corresponding FO expectations from \textsf{MadGraph5}~\cite{MG5_manual,MG5_BSM}. Specifically, we calculated the kinematic distributions of $H$, $W'$, and $Z'$ bosons accompanied by high-transverse momentum jets under various settings: a current LHC setup at a centre-of-mass energy of $\sqrt{s} = 13.6$ TeV, a future proton-proton collider at $\sqrt{s} = 100$ TeV, and a future $e^+e^-$ collider at $\sqrt{s} = 10$ TeV. We demonstrated that our BSM parton shower effectively captures the sequential BSM radiation under the collinear factorization approximation. Moreover, our simplified framework predicts that the behaviour of BSM events can be significantly altered by the emission of BSM bosons in future high-energy collider experiments, particularly when the mass of the BSM particle is relatively small. We intend to extend these generalized splitting functions to address coloured particles such as a dark shower~\cite{HiddenValley}, or extra-spin particles like gravitino or graviton in the next phase.

These novel parton shower schemes will be available with the \textsf{Herwig~7.4} public release.

\section*{Acknowledgments}

We would like to express our gratitude to our fellow \textsf{Herwig} authors for their invaluable contributions, with special thanks to N. Darvishi and P. Richardson for their insightful discussions and support. This work has received funding from the European Union's Horizon 2020 research and innovation programme as part of the Marie Skłodowska-Curie Innovative Training Network MCnetITN3 (grant agreement no. 722104) and from the UK Science and Technology Facilities Council (grant numbers~ST/T001011/1 and ST/T001038/1). This research was also supported in Korea by National Research Foundation grants NRF-2021R1A2C2014311, NRF-2008-00460 and the Promising-Pioneering Researcher Program through Seoul National University.

\newpage

\appendix
\section{BSM parton shower in \textsf{Herwig~7} interface}
\label{sec:manual}

\textsf{Herwig~7} is now equipped to handle all types of BSM boson showers, particularly those without colour structure or additional charges~\cite{herwig_manual,herwig7.0_release_note,herwig7.2_release_note}. The tutorial section titled \href{https://herwig.hepforge.org/tutorials/bsm/ufo.html}{``Using UFO Models"} provides comprehensive guidance on automatically setting up BSM features using any UFO model file~\cite{UFO_model_file}. Preparing \textsf{Herwig~7} for these simulations involves a few essential steps detailed in this documentation. First, users should allow the BSM shower as follows:
\mybox{
	\texttt{ufo2herwig <UFO\_directory> --enable-bsm-shower}\\
	\texttt{make}
}
\noindent To prevent any kind of disorder caused by including FCNC processes, this command automatically suppresses all FCNC-inducing splittings. One, however, can allow the inclusion of the FCNC processes by adding the ``\texttt{--allow-fcnc}" flag:
\mybox{
	\texttt{ufo2herwig <UFO\_directory>  --enable-bsm-shower --allow-fcnc}
}
\noindent Through \textsf{Herwig~7}'s \texttt{ufo2herwig} module, spin information, interaction types, coupling values, and others related to the parton shower process are written in a ``FRModel.model" file, which can be read by
\mybox{
	\texttt{read FRModel.model}
}
\noindent in any input file with the suffix ``.in". Note that if there are any changes in the UFO internal files, you should re-do
\mybox{
	\texttt{ufo2herwig <UFO\_directory> --enable-bsm-shower} \\
	\texttt{make}
}
\noindent to update \textsf{Herwig~7} modules, particularly the corresponding coupling values.

Inside the above-mentioned FRModel.model file, all BSM splittings are systematically defined. For example, if the model has a u-$\bar{\rm u}$-Zp vertex, a u\texttt{->}u,Zp splitting will be generated as
\mybox{
	\texttt{create Herwig::HalfHalfOneEWSplitFn uuZpSplitFnEW}
}
\noindent and
\mybox{
	\texttt{do /Herwig/Shower/SplittingGenerator:AddFinalSplitting u->u,Zp; uuZpSudakovEW}\\
	\texttt{u->u,Zp; uuZpSudakovEW}
}
\noindent which actually executes the radiation.

Generally speaking, only two types of coupling values are written in the model file:
\mybox{
	\texttt{cd /Herwig/FRModel/Particles}\\
	\texttt{set uuZpSplitFnEW:CouplingValue.Im <value>}\\
	\texttt{set uuZpSplitFnEW:CouplingValue.Re <value>}
}
\noindent where ``Im" and ``Re" are imaginary and real parts of the coupling value, respectively. However, for the spin-1/2 to spin-1/2 plus spin-1 splittings, left and right-handed couplings are considered separately 
\mybox{
	\texttt{cd /Herwig/FRModel/Particles}\\
	\texttt{set uuZpSplitFnEW:CouplingValue.Left.Im <value>}\\
	\texttt{set uuZpSplitFnEW:CouplingValue.Left.Re <value>}\\
	\texttt{set uuZpSplitFnEW:CouplingValue.Right.Im <value>}\\
	\texttt{set uuZpSplitFnEW:CouplingValue.Right.Re <value>}
}
\noindent In the same way, CP-even and -odd couplings in spin-1/2 to spin-1/2 plus spin-0 splittings are handled by
\mybox{
	\texttt{set bbh2SplitFnEW:CouplingValue.CP0.Im <value>}\\
	\texttt{set bbh2SplitFnEW:CouplingValue.CP0.Re <value>}\\
	\texttt{set bbh2SplitFnEW:CouplingValue.CP1.Im <value>}\\
	\texttt{set bbh2SplitFnEW:CouplingValue.CP1.Re <value>}
}
\noindent where CP0 (CP1) means CP-even (CP-odd) coupling. It is recommended that the user do not change the values directly in the FRModel.model file, but to do it in their input file.

As a final remark, BSM radiation is turned on automatically when the EW parton shower is switched, i.e.
\mybox{
	\texttt{set /Herwig/Shower/ShowerHandler:Interactions EWOnly}
}
\noindent or
\mybox{
	\texttt{set /Herwig/Shower/ShowerHandler:Interactions ALL}
}
\noindent where the first switches on only EW and BSM radiation but the second command does QED, QCD, EW, and BSM.

\bibliographystyle{JHEP}
\bibliography{references}

\providecommand{\href}[2]{#2}\begingroup\raggedright\begin{thebibliography}{10}

\bibitem{nu_oscillation_theory}
B.~{Pontecorvo}, \emph{{Neutrino Experiments and the Problem of Conservation of
  Leptonic Charge}}, {\emph{Soviet Journal of Experimental and Theoretical
  Physics} {\bfseries 26} (1968) 984}.

\bibitem{nu_oscillation_kamiokande}
{\scshape Super-Kamiokande} collaboration, \emph{Evidence for oscillation of
  atmospheric neutrinos},
  \href{https://doi.org/10.1103/PhysRevLett.81.1562}{\emph{Phys. Rev. Lett.}
  {\bfseries 81} (1998) 1562}.

\bibitem{nu_oscillation_SNO}
{\scshape SNO} collaboration, \emph{{Measurement of the rate of $\nu_e+d \to
  p+p+e^-$ interactions produced by $^8$B solar neutrinos at the Sudbury
  Neutrino Observatory}},
  \href{https://doi.org/10.1103/PhysRevLett.87.071301}{\emph{Phys. Rev. Lett.}
  {\bfseries 87} (2001) 071301}
  [\href{https://arxiv.org/abs/nucl-ex/0106015}{{\ttfamily nucl-ex/0106015}}].

\bibitem{w_mass_problem-cdf}
{\scshape CDF} collaboration, \emph{High-precision measurement of the
  \ensuremath{W} boson mass with the {CDF II} detector},
  \href{https://doi.org/10.1126/science.abk1781}{\emph{Science} {\bfseries 376}
  (2022) 170}
  [\href{https://arxiv.org/abs/https://www.science.org/doi/pdf/10.1126/science.abk1781}{{\ttfamily
  https://www.science.org/doi/pdf/10.1126/science.abk1781}}].

\bibitem{b_anomaly_summary}
S.~Bifani, S.~Descotes-Genon, A.R.~Vidal and M.-H.~Schune, \emph{Review of
  lepton universality tests in \ensuremath{B} decays},
  \href{https://doi.org/10.1088/1361-6471/aaf5de}{\emph{Journal of Physics G:
  Nuclear and Particle Physics} {\bfseries 46} (2018) 023001}.

\bibitem{muon_g-2_experiment}
{\scshape Muon $g\ensuremath{-}2$} collaboration, \emph{Measurement of the
  positive muon anomalous magnetic moment to 0.46 ppm},
  \href{https://doi.org/10.1103/PhysRevLett.126.141801}{\emph{Phys. Rev. Lett.}
  {\bfseries 126} (2021) 141801}.

\bibitem{herwig_manual}
M.~Bähr, S.~Gieseke, M.A.~Gigg, D.~Grellscheid, K.~Hamilton, O.~Latunde-Dada
  et~al., \emph{{Herwig++} physics and manual},
  \href{https://doi.org/10.1140/epjc/s10052-008-0798-9}{\emph{The European
  Physical Journal C} {\bfseries 58} (2008) 639}.

\bibitem{herwig7.0_release_note}
J.~Bellm, S.~Gieseke, D.~Grellscheid, S.~Plätzer, M.~Rauch, C.~Reuschle
  et~al., \emph{Herwig 7.0/{Herwig++} 3.0 release note},
  \href{https://doi.org/10.1140/epjc/s10052-016-4018-8}{\emph{The European
  Physical Journal C} {\bfseries 76} (2016) }.

\bibitem{herwig7.2_release_note}
J.~Bellm, G.~Bewick, S.F.~Ravasio, S.~Gieseke, D.~Grellscheid,
  P.~Kirchgae{\ss}er et~al., \emph{Herwig 7.2 release note},
  \href{https://doi.org/10.1140/epjc/s10052-020-8011-x}{\emph{The European
  Physical Journal C} {\bfseries 80} (2020) }.

\bibitem{Bewick:2023tfi}
G.~Bewick et~al., \emph{{Herwig 7.3 Release Note}},
  \href{https://arxiv.org/abs/2312.05175}{{\ttfamily 2312.05175}}.

\bibitem{Masouminia:2023zhb}
M.R.~Masouminia and P.~Richardson, \emph{{Hadronization and Decay of Excited
  Heavy Hadrons in Herwig 7}},
  \href{https://arxiv.org/abs/2312.02757}{{\ttfamily 2312.02757}}.

\bibitem{pythia_manual}
C.~Bierlich, S.~Chakraborty, N.~Desai, L.~Gellersen, I.~Helenius, P.~Ilten
  et~al., \emph{A comprehensive guide to the physics and usage of {PYTHIA}
  8.3},  \href{https://arxiv.org/abs/2203.11601}{{\ttfamily 2203.11601}}.

\bibitem{Sherpa1.1_manual}
T.~Gleisberg, S.~Höche, F.~Krauss, M.~Schönherr, S.~Schumann, F.~Siegert
  et~al., \emph{Event generation with {SHERPA} 1.1},
  \href{https://doi.org/10.1088/1126-6708/2009/02/007}{\emph{Journal of High
  Energy Physics} {\bfseries 2009} (2009) 007}.

\bibitem{Sherpa2.2_manual}
E.~Bothmann, G.S.~Chahal, S.~Höche, J.~Krause, F.~Krauss, S.~Kuttimalai
  et~al., \emph{{Event generation with Sherpa 2.2}},
  \href{https://doi.org/10.21468/SciPostPhys.7.3.034}{\emph{SciPost Phys.}
  {\bfseries 7} (2019) 034}.

\bibitem{Sherpa_BSM}
S.~Höche, S.~Kuttimalai, S.~Schumann and F.~Siegert, \emph{Beyond standard
  model calculations with {Sherpa}},
  \href{https://doi.org/10.1140/epjc/s10052-015-3338-4}{\emph{The European
  Physical Journal C} {\bfseries 75} (2015) }.

\bibitem{MG5_manual}
J.~Alwall, R.~Frederix, S.~Frixione, V.~Hirschi, F.~Maltoni, O.~Mattelaer
  et~al., \emph{The automated computation of tree-level and next-to-leading
  order differential cross sections, and their matching to parton shower
  simulations}, \href{https://doi.org/10.1007/jhep07(2014)079}{\emph{Journal of
  High Energy Physics} {\bfseries 2014} (2014) 079}.

\bibitem{MG5_BSM}
S.~Frixione, B.~Fuks, V.~Hirschi, K.~Mawatari, H.-S.~Shao, P.A.~Sunder
  et~al.\href{https://doi.org/10.1007/JHEP12(2019)008}{\emph{JHEP} {\bfseries
  12} (2019) 008} [\href{https://arxiv.org/abs/1907.04898}{{\ttfamily
  1907.04898}}].

\bibitem{neccessity_of_ewk_radiation}
W.~Beenakker and A.~Werthenbach, \emph{New insights into the perturbative
  structure of electroweak {Sudakov} logarithms},
  \href{https://doi.org/https://doi.org/10.1016/S0370-2693(00)00900-X}{\emph{Physics
  Letters B} {\bfseries 489} (2000) 148}.

\bibitem{Darvishi:2020paz}
N.~Darvishi and M.R.~Masouminia, \emph{{Signature of the Maximally Symmetric
  2HDM via $W^{\pm}/Z$-Quadruplet Productions at the LHC}},
  \href{https://doi.org/10.1103/PhysRevD.103.095031}{\emph{Phys. Rev. D}
  {\bfseries 103} (2021) 095031}
  [\href{https://arxiv.org/abs/2012.14746}{{\ttfamily 2012.14746}}].

\bibitem{pythia_ewk_radiation}
J.R.~Christiansen and T.~Sjöstrand, \emph{Weak gauge boson radiation in parton
  showers}, \href{https://doi.org/10.1007/jhep04(2014)115}{\emph{Journal of
  High Energy Physics} {\bfseries 2014} (2014) 115}.

\bibitem{herwig_ewk_radiation}
M.R.~Masouminia and P.~Richardson, \emph{Implementation of angularly ordered
  electroweak parton shower in {Herwig} 7},
  \href{https://doi.org/10.1007/jhep04(2022)112}{\emph{Journal of High Energy
  Physics} {\bfseries 2022} (2022) 112}.

\bibitem{vincia_ewk_radiation}
R.~Kleiss and R.~Verheyen, \emph{Collinear electroweak radiation in antenna
  parton showers},
  \href{https://doi.org/10.1140/epjc/s10052-020-08510-w}{\emph{The European
  Physical Journal C} {\bfseries 80} (2020) }.

\bibitem{ewk_radiation-theory_GET}
J.~Chen, \emph{{On the Feynman Rules of Massive Gauge Theory in Physical
  Gauges}},  \href{https://arxiv.org/abs/1902.06738}{{\ttfamily 1902.06738}}.

\bibitem{CMS-low_mass_Z'-1}
{\scshape CMS} collaboration, \emph{{Search for a Narrow Resonance Lighter than
  200 GeV Decaying to a Pair of Muons in Proton-Proton Collisions at $\sqrt{s}$
  = 13 TeV}}, \href{https://doi.org/10.1103/PhysRevLett.124.131802}{\emph{Phys.
  Rev. Lett.} {\bfseries 124} (2020) 131802}.

\bibitem{CMS-low_mass_Z'-2}
{\scshape CMS} collaboration, \emph{{Search for long-lived particles decaying
  into muon pairs in proton-proton collisions at $ \sqrt{s} $ = 13 TeV
  collected with a dedicated high-rate data stream}},
  \href{https://doi.org/10.1007/JHEP04(2022)062}{\emph{JHEP} {\bfseries 04}
  (2022) 062} [\href{https://arxiv.org/abs/2112.13769}{{\ttfamily
  2112.13769}}].

\bibitem{CMS-low_mass_Z'-3}
{\scshape CMS} collaboration, \emph{Search for long-lived particles decaying to
  a pair of muons in proton-proton collisions at $\sqrt{s}= 13$ tev},
  \href{https://doi.org/10.1007/jhep05(2023)228}{\emph{Journal of High Energy
  Physics} {\bfseries 2023} (2023) 228}.

\bibitem{LHCb-low_mass_Z'}
{\scshape LHCb} collaboration, \emph{{Searches for low-mass dimuon
  resonances}}, \href{https://doi.org/10.1007/JHEP10(2020)156}{\emph{JHEP}
  {\bfseries 10} (2020) 156}
  [\href{https://arxiv.org/abs/2007.03923}{{\ttfamily 2007.03923}}].

\bibitem{intro_to_AO}
{G. Marchesini and B.R. Webber}, \emph{{Simulation of QCD jets including soft
  gluon interference}},
  \href{https://doi.org/https://doi.org/10.1016/0550-3213(84)90463-2}{\emph{Nuclear
  Physics B} {\bfseries 238} (1984) 1}.

\bibitem{feynrules2_manual}
A.~Alloul, N.D.~Christensen, C.~Degrande, C.~Duhr and B.~Fuks, \emph{{FeynRules
  2.0 -- A complete toolbox for tree-level phenomenology}},
  \href{https://doi.org/10.1016/j.cpc.2014.04.012}{\emph{Comput. Phys. Commun.}
  {\bfseries 185} (2014) 2250}
  [\href{https://arxiv.org/abs/1310.1921}{{\ttfamily 1310.1921}}].

\bibitem{UFO_model_file}
C.~Degrande, C.~Duhr, B.~Fuks, D.~Grellscheid, O.~Mattelaer and T.~Reiter,
  \emph{{UFO} {\textendash} the universal {FeynRules} output},
  \href{https://doi.org/10.1016/j.cpc.2012.01.022}{\emph{Computer Physics
  Communications} {\bfseries 183} (2012) 1201}.

\bibitem{Bewick:2019rbu}
G.~Bewick, S.~Ferrario~Ravasio, P.~Richardson and M.H.~Seymour,
  \emph{{Logarithmic accuracy of angular-ordered parton showers}},
  \href{https://doi.org/10.1007/JHEP04(2020)019}{\emph{JHEP} {\bfseries 04}
  (2020) 019} [\href{https://arxiv.org/abs/1904.11866}{{\ttfamily
  1904.11866}}].

\bibitem{Bewick:2021nhc}
G.~Bewick, S.~Ferrario~Ravasio, P.~Richardson and M.H.~Seymour, \emph{{Initial
  state radiation in the Herwig 7 angular-ordered parton shower}},
  \href{https://doi.org/10.1007/JHEP01(2022)026}{\emph{JHEP} {\bfseries 01}
  (2022) 026} [\href{https://arxiv.org/abs/2107.04051}{{\ttfamily
  2107.04051}}].

\bibitem{Hoche:2015sya}
S.~H\"oche and S.~Prestel, \emph{{The midpoint between dipole and parton
  showers}}, \href{https://doi.org/10.1140/epjc/s10052-015-3684-2}{\emph{Eur.
  Phys. J. C} {\bfseries 75} (2015) 461}
  [\href{https://arxiv.org/abs/1506.05057}{{\ttfamily 1506.05057}}].

\bibitem{Dasgupta:2020fwr}
M.~Dasgupta, F.A.~Dreyer, K.~Hamilton, P.F.~Monni, G.P.~Salam and G.~Soyez,
  \emph{{Parton showers beyond leading logarithmic accuracy}},
  \href{https://doi.org/10.1103/PhysRevLett.125.052002}{\emph{Phys. Rev. Lett.}
  {\bfseries 125} (2020) 052002}
  [\href{https://arxiv.org/abs/2002.11114}{{\ttfamily 2002.11114}}].

\bibitem{Nagy:2020rmk}
Z.~Nagy and D.E.~Soper, \emph{{Summations of large logarithms by parton
  showers}}, \href{https://doi.org/10.1103/PhysRevD.104.054049}{\emph{Phys.
  Rev. D} {\bfseries 104} (2021) 054049}
  [\href{https://arxiv.org/abs/2011.04773}{{\ttfamily 2011.04773}}].

\bibitem{Dawson}
S.~Dawson, \emph{{The Effective W Approximation}},
  \href{https://doi.org/10.1016/0550-3213(85)90038-0}{\emph{Nucl. Phys. B}
  {\bfseries 249} (1985) 42}.

\bibitem{SUSY_splitting-2}
S.~Catani, S.~Dittmaier and Z.~Tr{\'{o}}cs{\'{a}}nyi, \emph{One-loop singular
  behaviour of {QCD} and {SUSY} {QCD} amplitudes with massive partons},
  \href{https://doi.org/10.1016/s0370-2693(01)00065-x}{\emph{Physics Letters B}
  {\bfseries 500} (2001) 149}.

\bibitem{Altarelli_PS_introduction}
G.~Altarelli and G.~Parisi, \emph{{Asymptotic Freedom in Parton Language}},
  \href{https://doi.org/10.1016/0550-3213(77)90384-4}{\emph{Nucl. Phys. B}
  {\bfseries 126} (1977) 298}.

\bibitem{2_loop_sf}
{S.D. Badger and E.W.N. Glover}, \emph{Two-loop splitting functions in {QCD}},
  \href{https://doi.org/10.1088/1126-6708/2004/07/040}{\emph{Journal of High
  Energy Physics} {\bfseries 2004} (2004) 040}.

\bibitem{mhv_rules}
{T.G. Birthwright and E.W.N. Glover and V.V. Khoze and P. Marquard},
  \emph{Collinear limits in {QCD} from {MHV} rules},
  \href{https://doi.org/10.1088/1126-6708/2005/07/068}{\emph{Journal of High
  Energy Physics} {\bfseries 2005} (2005) 068}.

\bibitem{2hdm-ufo}
C.~Degrande, \emph{Automatic evaluation of {UV} and \ensuremath{R2} terms for
  beyond the {Standard Model Lagrangians}: A proof-of-principle},
  \href{https://doi.org/10.1016/j.cpc.2015.08.015}{\emph{Computer Physics
  Communications} {\bfseries 197} (2015) 239}.

\bibitem{Darvishi:2019ltl}
N.~Darvishi and A.~Pilaftsis, \emph{{Quartic Coupling Unification in the
  Maximally Symmetric 2HDM}},
  \href{https://doi.org/10.1103/PhysRevD.99.115014}{\emph{Phys. Rev. D}
  {\bfseries 99} (2019) 115014}
  [\href{https://arxiv.org/abs/1904.06723}{{\ttfamily 1904.06723}}].

\bibitem{Darvishi:2023nft}
N.~Darvishi, A.~Pilaftsis and J.-H.~Yu, \emph{{Maximising CP Violation in
  Naturally Aligned Two-Higgs Doublet Models}},
  \href{https://arxiv.org/abs/2312.00882}{{\ttfamily 2312.00882}}.

\bibitem{bl4-ufo-1}
S.~Amrith, J.M.~Butterworth, F.F.~Deppisch, W.~Liu, A.~Varma and D.~Yallup,
  \emph{{LHC} constraints on a {B -- L} gauge model using {Contur}},
  \href{https://doi.org/10.1007/jhep05(2019)154}{\emph{Journal of High Energy
  Physics} {\bfseries 2019} (2019) 154}.

\bibitem{bl4-ufo-2}
F.F.~Deppisch, W.~Liu and M.~Mitra, \emph{Long-lived heavy neutrinos from
  {Higgs} decays}, \href{https://doi.org/10.1007/jhep08(2018)181}{\emph{Journal
  of High Energy Physics} {\bfseries 2018} (2018) 181}.

\bibitem{bl4-ufo-3}
L.~Basso, A.~Belyaev, S.~Moretti and C.H.~Shepherd-Themistocleous,
  \emph{Phenomenology of the minimal {B -- L} extension of the {Standard
  Model}: \ensuremath{Z'} and neutrinos},
  \href{https://doi.org/10.1103/physrevd.80.055030}{\emph{Physical Review D}
  {\bfseries 80} (2009) }.

\bibitem{effW-ufo-1}
Z.~Sullivan, \emph{Fully differential \ensuremath{W'} production and decay at
  {Next-to-Leading Order} in {QCD}},
  \href{https://doi.org/10.1103/physrevd.66.075011}{\emph{Physical Review D}
  {\bfseries 66} (2002) }.

\bibitem{effW-ufo-2}
D.~Duffty and Z.~Sullivan, \emph{Model independent reach for {W-prime} bosons
  at the {LHC}},
  \href{https://doi.org/10.1103/physrevd.86.075018}{\emph{Physical Review D}
  {\bfseries 86} (2012) }.

\bibitem{heavyquark}
{G. Marchesini and B.R. Webber}, \emph{Simulation of {QCD} coherence in heavy
  quark production and decay},
  \href{https://doi.org/https://doi.org/10.1016/0550-3213(90)90310-A}{\emph{Nuclear
  Physics B} {\bfseries 330} (1990) 261}.

\bibitem{Gieseke:2003rz}
S.~Gieseke, P.~Stephens and B.~Webber, \emph{{New formalism for QCD parton
  showers}}, \href{https://doi.org/10.1088/1126-6708/2003/12/045}{\emph{JHEP}
  {\bfseries 12} (2003) 045}
  [\href{https://arxiv.org/abs/hep-ph/0310083}{{\ttfamily hep-ph/0310083}}].

\bibitem{Gieseke:2003hm}
S.~Gieseke, A.~Ribon, M.H.~Seymour, P.~Stephens and B.~Webber,
  \emph{{{Herwig++} 1.0: An Event generator for \ensuremath{e^+ e^-}
  annihilation}},
  \href{https://doi.org/10.1088/1126-6708/2004/02/005}{\emph{JHEP} {\bfseries
  02} (2004) 005} [\href{https://arxiv.org/abs/hep-ph/0311208}{{\ttfamily
  hep-ph/0311208}}].

\bibitem{Reichelt:2017hts}
D.~Reichelt, P.~Richardson and A.~Siodmok, \emph{{Improving the Simulation of
  Quark and Gluon Jets with Herwig 7}},
  \href{https://doi.org/10.1140/epjc/s10052-017-5374-8}{\emph{Eur. Phys. J. C}
  {\bfseries 77} (2017) 876}
  [\href{https://arxiv.org/abs/1708.01491}{{\ttfamily 1708.01491}}].

\bibitem{herwig7.1_release_note}
J.~Bellm et~al., \emph{{Herwig 7.1 Release Note}},
  \href{https://arxiv.org/abs/1705.06919}{{\ttfamily 1705.06919}}.

\bibitem{SUSY_splitting}
W.~Beenakker, R.~Höpker, M.~Spira and P.~Zerwas, \emph{Squark and gluino
  production at hadron colliders},
  \href{https://doi.org/10.1016/s0550-3213(97)80027-2}{\emph{Nuclear Physics B}
  {\bfseries 492} (1997) 51}.

\bibitem{2HDM_THC-1}
F.~Arco, S.~Heinemeyer and M.J.~Herrero, \emph{Exploring sizable triple {Higgs}
  couplings in the {2HDM}},
  \href{https://doi.org/10.1140/epjc/s10052-020-8406-8}{\emph{The European
  Physical Journal C} {\bfseries 80} (2020) }.

\bibitem{2HDM_THC-2}
F.~Arco, S.~Heinemeyer and M.J.~Herrero, \emph{Triple {Higgs} couplings in the
  {2HDM}: the complete picture},
  \href{https://doi.org/10.1140/epjc/s10052-022-10485-9}{\emph{The European
  Physical Journal C} {\bfseries 82} (2022) }.

\bibitem{SUSYhiggs}
J.F.~Gunion and H.E.~Haber, \emph{Higgs bosons in supersymmetric models {(I)}},
  \href{https://doi.org/https://doi.org/10.1016/0550-3213(86)90340-8}{\emph{Nuclear
  Physics B} {\bfseries 272} (1986) 1}.

\bibitem{Pilaftsis:1999qt}
A.~Pilaftsis and C.E.M.~Wagner, \emph{{Higgs bosons in the minimal
  supersymmetric standard model with explicit CP violation}},
  \href{https://doi.org/10.1016/S0550-3213(99)00261-8}{\emph{Nucl. Phys. B}
  {\bfseries 553} (1999) 3}
  [\href{https://arxiv.org/abs/hep-ph/9902371}{{\ttfamily hep-ph/9902371}}].

\bibitem{bible}
{Yu.L. Dokshitzer}, \emph{Perturbative {QCD} for beginners},  2002.

\bibitem{DLY-1}
S.D.~Drell, D.J.~Levy and T.-M.~Yan, \emph{Theory of {Deep-Inelastic
  Lepton-Nucleon Scattering} and {Lepton-Pair Annihilation Processes}. {I}},
  \href{https://doi.org/10.1103/PhysRev.187.2159}{\emph{Phys. Rev.} {\bfseries
  187} (1969) 2159}.

\bibitem{DLY-2}
S.D.~Drell, D.J.~Levy and T.-M.~Yan, \emph{Theory of {Deep-Inelastic
  Lepton-Nucleon Scattering} and {Lepton-Pair Annihilation Processes}. {III.
  Deep-Inelastic Electron-Positron Annihilation}},
  \href{https://doi.org/10.1103/PhysRevD.1.1617}{\emph{Phys. Rev. D} {\bfseries
  1} (1970) 1617}.

\bibitem{DLY-NLO}
{J Blümlein and V Ravindran and W.L van Neerven}, \emph{On the
  {Drell{\textendash}Levy{\textendash}Yan} relation to
  \ensuremath{O(\alpha_s^2)}},
  \href{https://doi.org/10.1016/s0550-3213(00)00422-3}{\emph{Nuclear Physics B}
  {\bfseries 586} (2000) 349}.

\bibitem{HiddenValley}
L.~Carloni and T.~Sjöstrand, \emph{Visible effects of invisible hidden valley
  radiation}, \href{https://doi.org/10.1007/jhep09(2010)105}{\emph{Journal of
  High Energy Physics} {\bfseries 2010} (2010) 105}.

\end{thebibliography}\endgroup


\clearpage
\begin{figure}
\centering
    \begin{subfigure}[b]{0.49\textwidth}
        \centering
        \includegraphics[width=\textwidth]{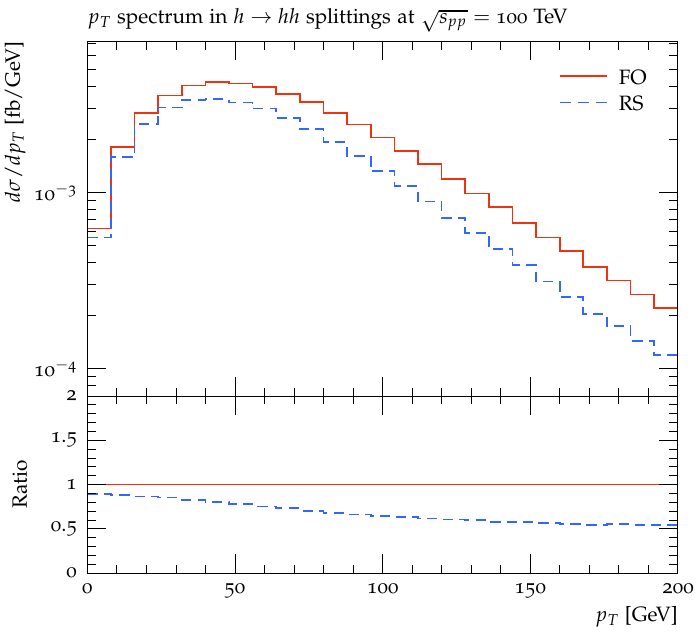}
        \caption{}
        \label{f:zhh_pt}
    \end{subfigure}
    \begin{subfigure}[b]{0.49\textwidth}
        \centering
        \includegraphics[width=\textwidth]{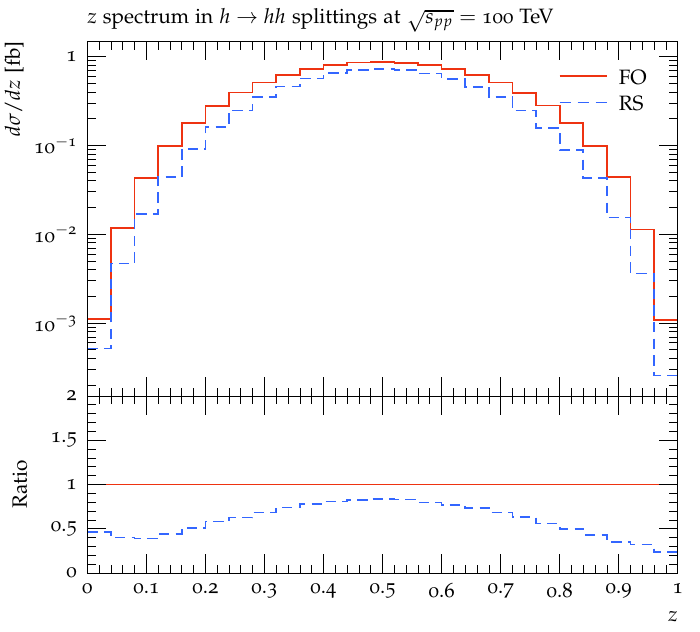}
        \caption{}
        \label{f:zhh_z}
    \end{subfigure}
    \begin{subfigure}[b]{0.49\textwidth}
        \centering
        \includegraphics[width=\textwidth]{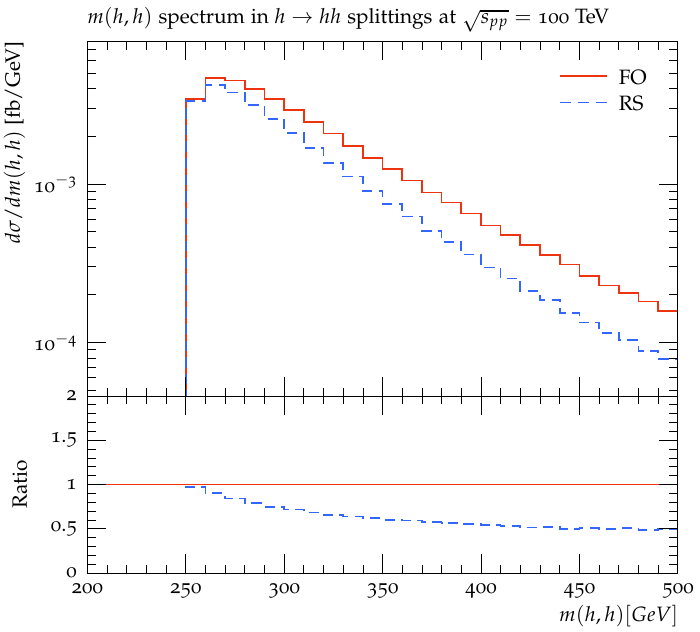}
        \caption{}
        \label{f:zhh_mij}
    \end{subfigure}
    \begin{subfigure}[b]{0.49\textwidth}
        \centering
        \includegraphics[width=\textwidth]{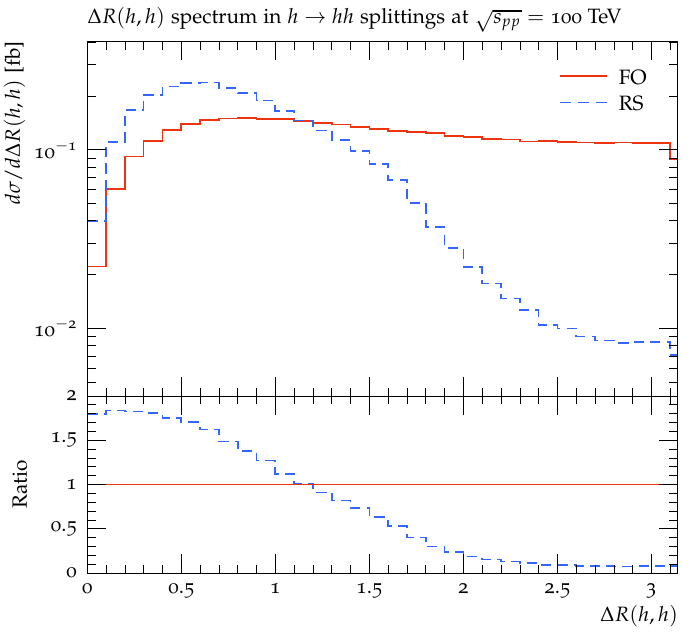}
        \caption{}
        \label{f:zhh_drij}
    \end{subfigure}
    \caption{Performance test for the SM $h\rightarrow hh$ splitting in \textsf{Herwig~7} for $\sqrt{s} =$ 100 TeV. \ref{f:zhh_pt}, \ref{f:zhh_z}, \ref{f:zhh_mij}, and \ref{f:zhh_drij} show the differential rate of $h$ boson radiation as functions of the transverse momentum, the light-cone momentum, the mass of the di-Higgs system, and the $\Delta R$ between the two emitted Higgs bosons respectively in resummed EW (blue) and FO calculations (red).}
    \label{f:zhh}
\end{figure} 

\begin{figure} [!h]
\centering
    \begin{subfigure}[b]{0.49\textwidth}
        \centering
        \includegraphics[width=\textwidth]{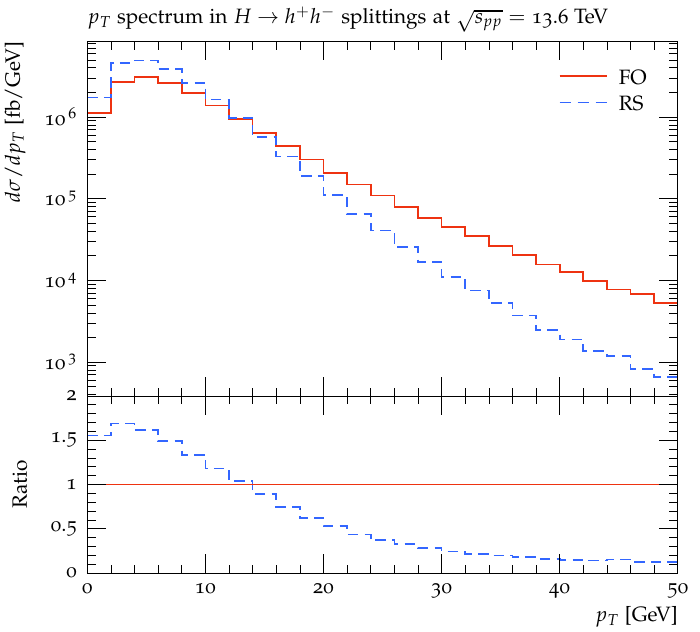}
        \caption{}
        \label{f:hmhp_pt}
    \end{subfigure}
    \begin{subfigure}[b]{0.49\textwidth}
        \centering
        \includegraphics[width=\textwidth]{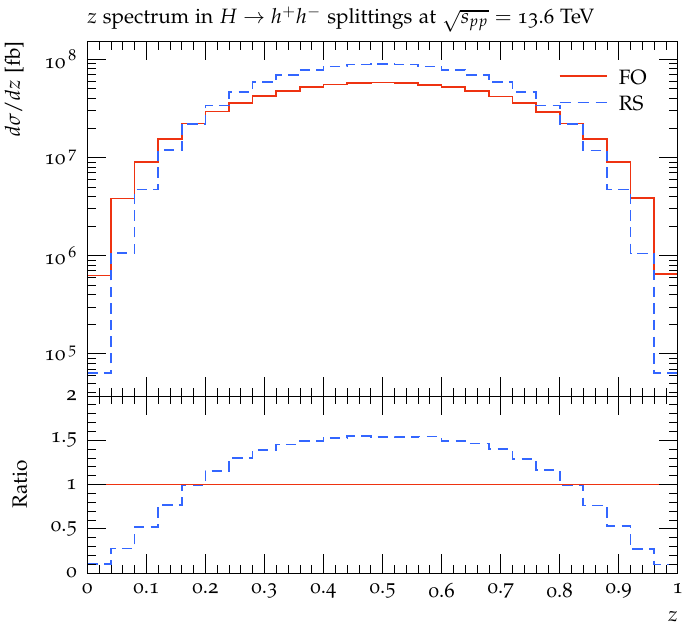}
        \caption{}
        \label{f:hmhp_z}
    \end{subfigure}
    \begin{subfigure}[b]{0.49\textwidth}
        \centering
        \includegraphics[width=\textwidth]{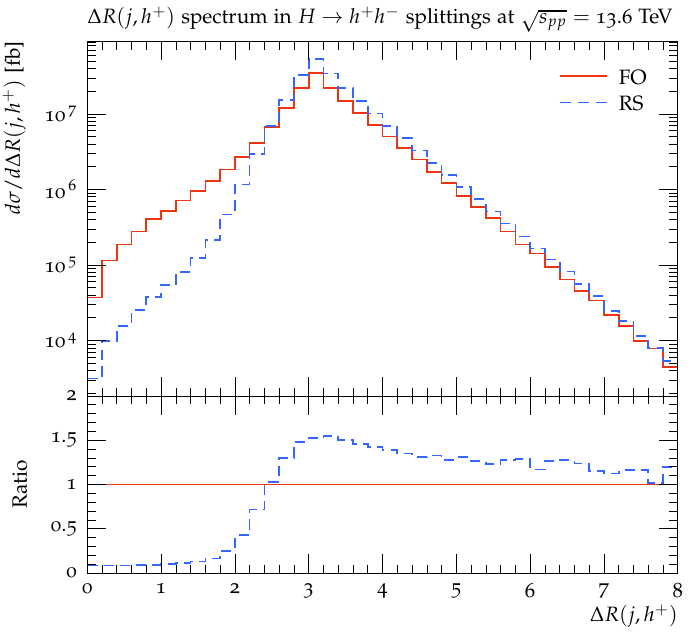}
        \caption{}
        \label{f:hmhp_drri}
    \end{subfigure}
    \begin{subfigure}[b]{0.49\textwidth}
        \centering
        \includegraphics[width=\textwidth]{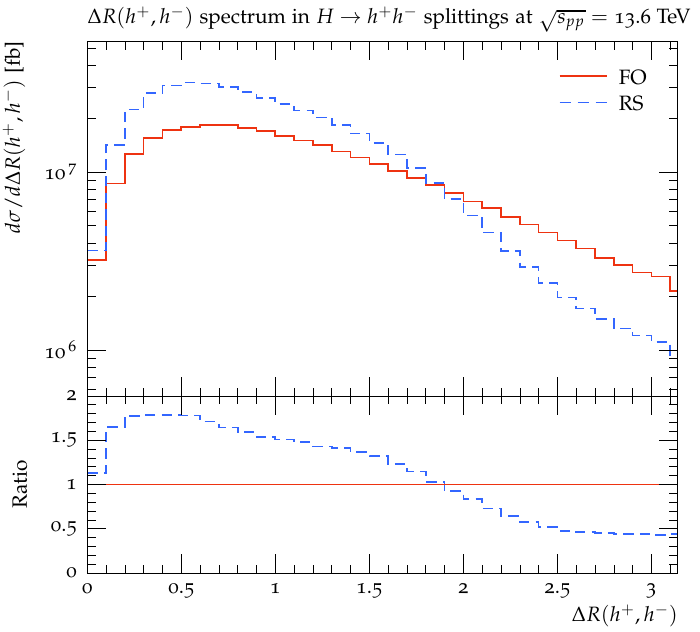}
        \caption{}
        \label{f:hmhp_drij}
    \end{subfigure}
    \caption{Performance test for $H \to h^+ h^-$ branching in \textsf{Herwig~7} for $\sqrt{s} = 13.6$ TeV. \ref{f:hmhp_pt}, \ref{f:hmhp_z}, \ref{f:hmhp_drri}, and \ref{f:hmhp_drij} show the differential rate of charged Higgs boson radiation as functions of the transverse momentum, the light-cone momentum, the $\Delta R$ between the recoiled particle and the emitted charged Higgs boson, and the $\Delta R$ between two emitted Higgs bosons respectively. $H$ and $h^\pm$ masses are both 10 GeV.}
    \label{f:hmhp}
\end{figure}

\begin{figure} [h!]
\centering
    \begin{subfigure}[b]{0.49\textwidth}
        \centering
        \includegraphics[width=\textwidth]{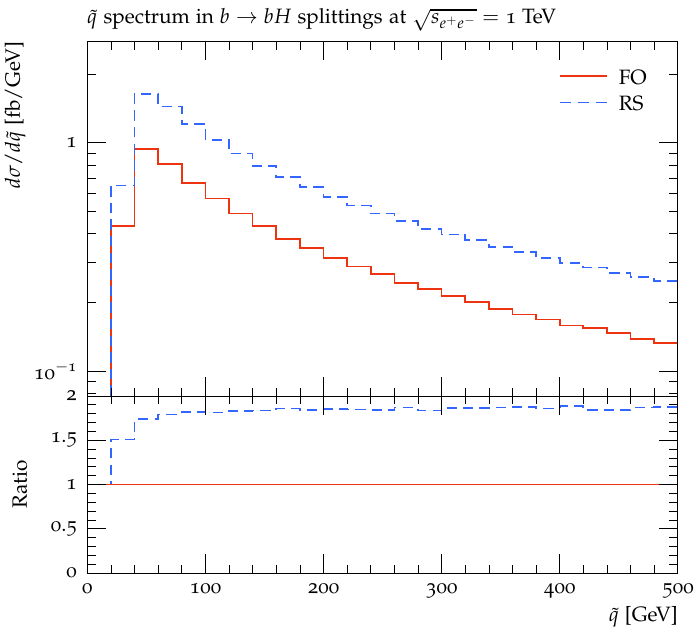}
        \caption{}
        \label{f:bbh2 mh2-10 qt}
    \end{subfigure}
    \begin{subfigure}[b]{0.49\textwidth}
        \centering
        \includegraphics[width=\textwidth]{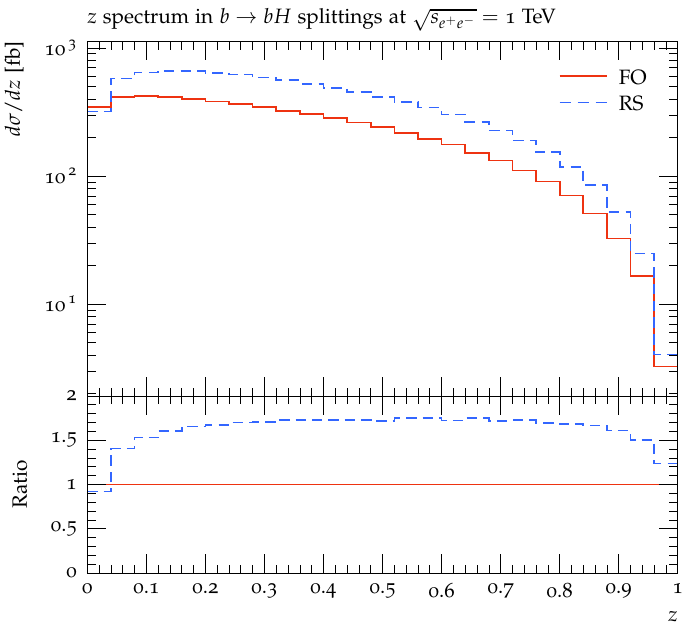}
        \caption{}
        \label{f:bbh2 mh2-10 z}
    \end{subfigure}
    \begin{subfigure}[b]{0.49\textwidth}
        \centering
        \includegraphics[width=\textwidth]{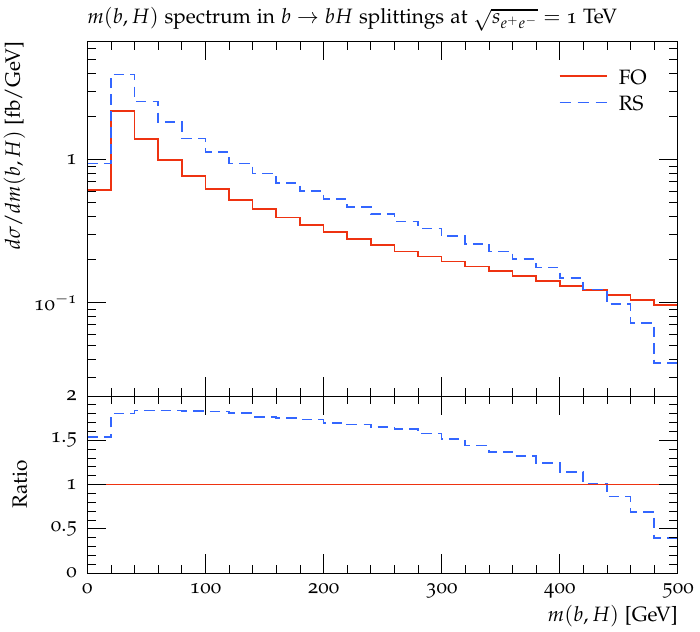}
        \caption{}
        \label{f:bbh2 mh2-10 m}
    \end{subfigure}
    \begin{subfigure}[b]{0.49\textwidth}
        \centering
        \includegraphics[width=\textwidth]{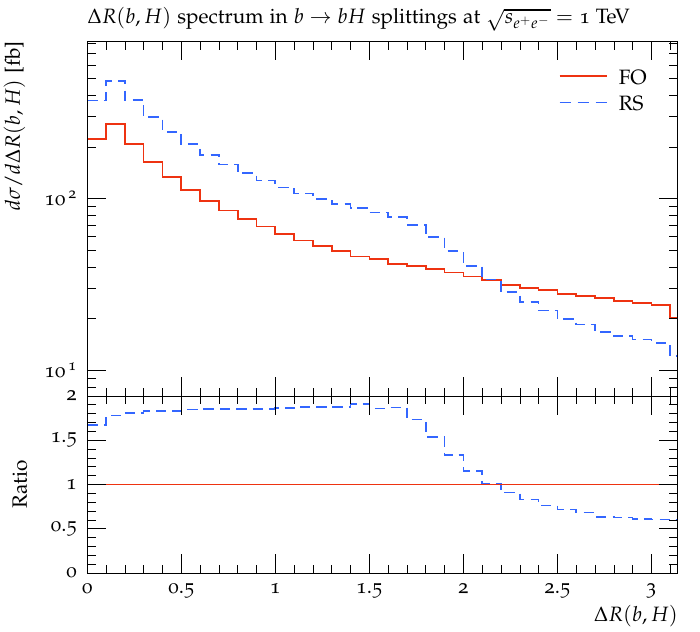}
        \caption{}
        \label{f:bbh2 mh2-10 drij}
    \end{subfigure}
    \caption{Performance test for $b(\Bar{b})\rightarrow b(\Bar{b}) H$ branching in \textsf{Herwig~7} with the 1 TeV $e^+ e^-$ collision setup, where the mass of the $H$ boson is assumed as 10 GeV. \ref{f:bbh2 mh2-10 qt}, \ref{f:bbh2 mh2-10 z}, \ref{f:bbh2 mh2-10 m}, and \ref{f:bbh2 mh2-10 drij} show the differential rate of BSM $H$ boson radiation as functions of the $\Tilde{q}$, the light-cone momentum, and mass and $\Delta R$ between two emitted particles respectively.}
    \label{f:bbh2 mh2-10}
\end{figure} 

\begin{figure} [h!]
\centering
    \begin{subfigure}[b]{0.49\textwidth}
        \centering
        \includegraphics[width=\textwidth]{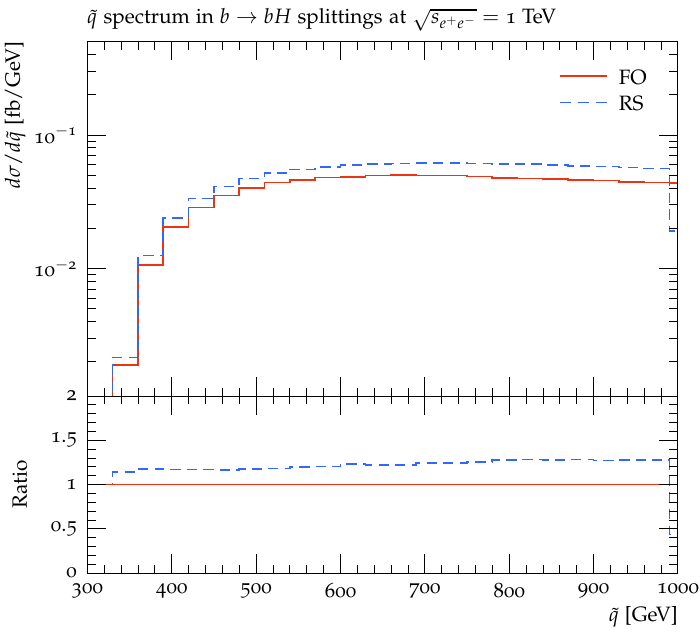}
        \caption{}
        \label{f:bbh2 mh2-130 qt}
    \end{subfigure}
    \begin{subfigure}[b]{0.49\textwidth}
        \centering
        \includegraphics[width=\textwidth]{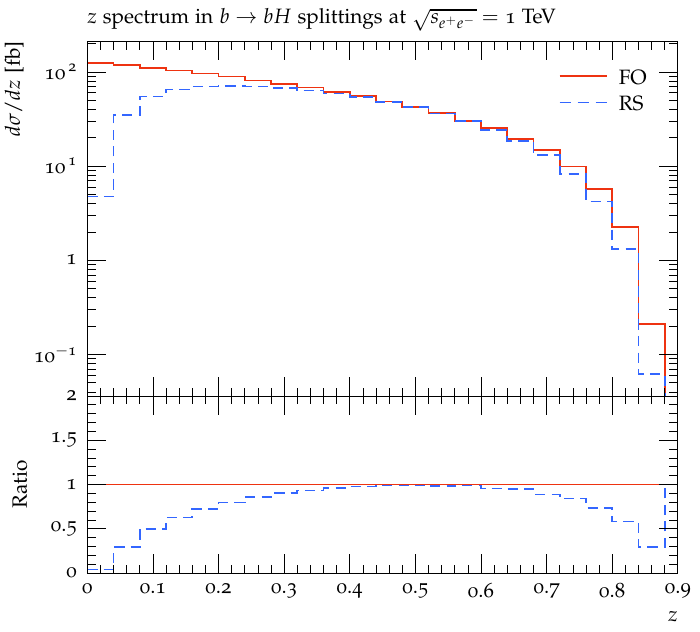}
        \caption{}
        \label{f:bbh2 mh2-130 z}
    \end{subfigure}
    \begin{subfigure}[b]{0.49\textwidth}
        \centering
        \includegraphics[width=\textwidth]{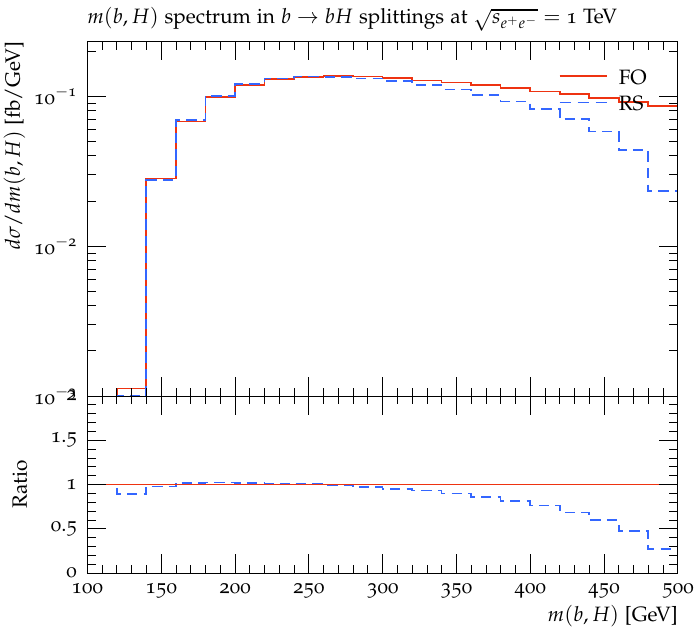}
        \caption{}
        \label{f:bbh2 mh2-130 drri}
    \end{subfigure}
    \begin{subfigure}[b]{0.49\textwidth}
        \centering
        \includegraphics[width=\textwidth]{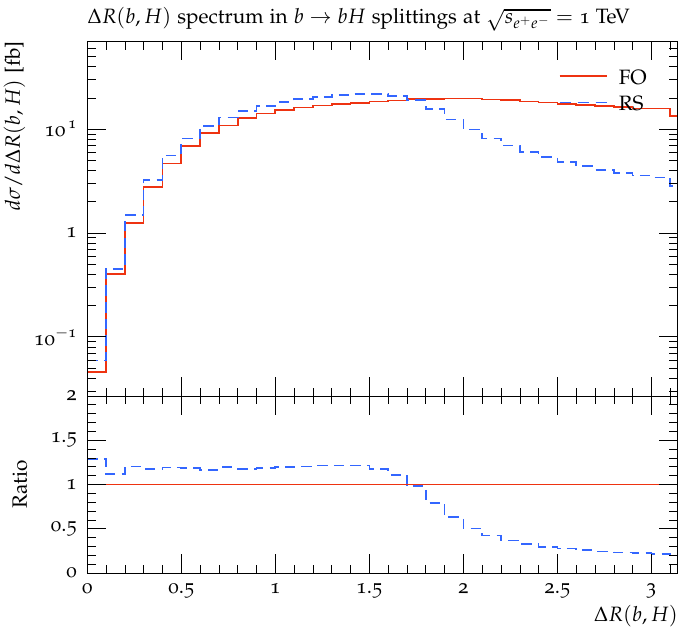}
        \caption{}
        \label{f:bbh2 mh2-130 mij}
    \end{subfigure}
    \caption{Performance test for $b(\Bar{b})\rightarrow b(\Bar{b}) H$ branching in \textsf{Herwig~7}, where the mass of the $H$ boson is assumed as 130 GeV. The notation of the figure is the same as in figure~\ref{f:bbh2 mh2-10}.}
    \label{f:bbh2 mh2-130}
\end{figure} 

\begin{figure} [h!]
\centering
    \begin{subfigure}[b]{0.49\textwidth}
        \centering
        \includegraphics[width=\textwidth]{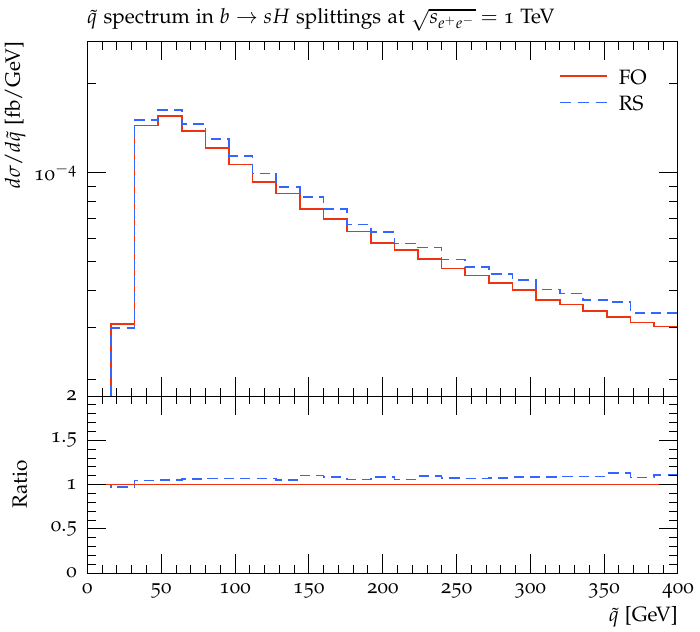}
        \caption{}
        \label{f:bsh2 mh2-10 qt}
    \end{subfigure}
    \begin{subfigure}[b]{0.49\textwidth}
        \centering
        \includegraphics[width=\textwidth]{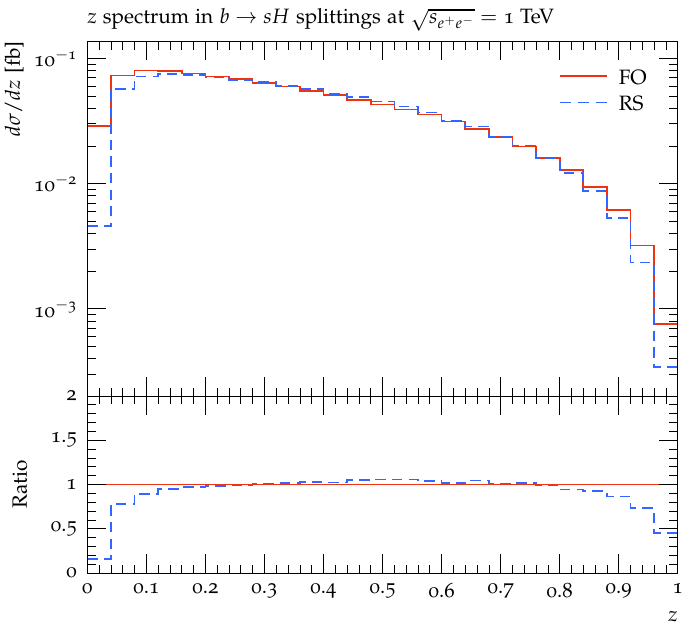}
        \caption{}
        \label{f:bsh2 mh2-10 z}
    \end{subfigure}
    \begin{subfigure}[b]{0.49\textwidth}
        \centering
        \includegraphics[width=\textwidth]{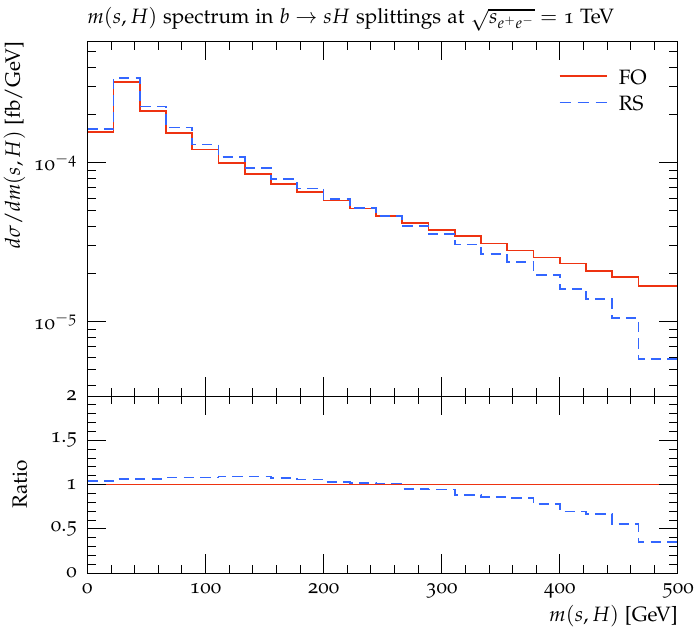}
        \caption{}
        \label{f:bsh2 mh2-10 drri}
    \end{subfigure}
    \begin{subfigure}[b]{0.49\textwidth}
        \centering
        \includegraphics[width=\textwidth]{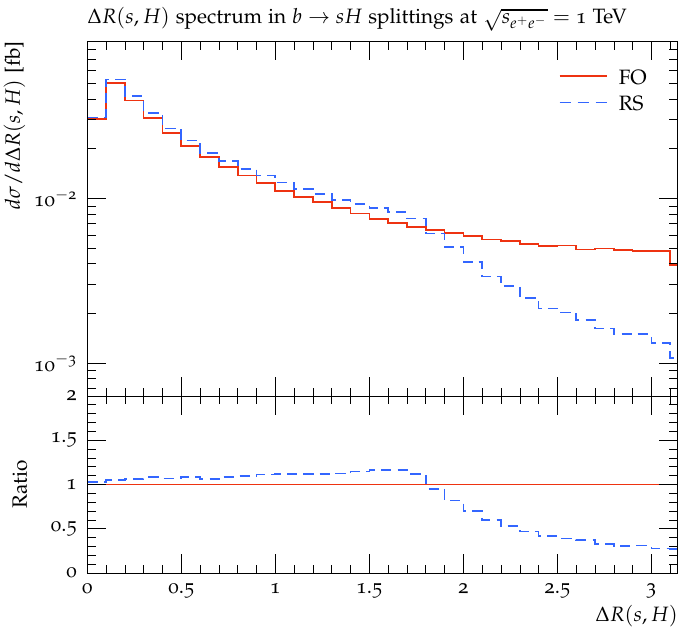}
        \caption{}
        \label{f:bsh2 mh2-10 mij}
    \end{subfigure}
    \caption{Performance test for $b(\Bar{b})\rightarrow s(\Bar{s}) H$ branching in \textsf{Herwig~7}, where the mass of the $H$ boson is assumed as 10 GeV. The notation of the figure is the same as in figure~\ref{f:bbh2 mh2-10}.}
    \label{f:bsh2 mh2-10}
\end{figure} 

\begin{figure} [h!]
\centering
    \begin{subfigure}[b]{0.49\textwidth}
        \centering
        \includegraphics[width=\textwidth]{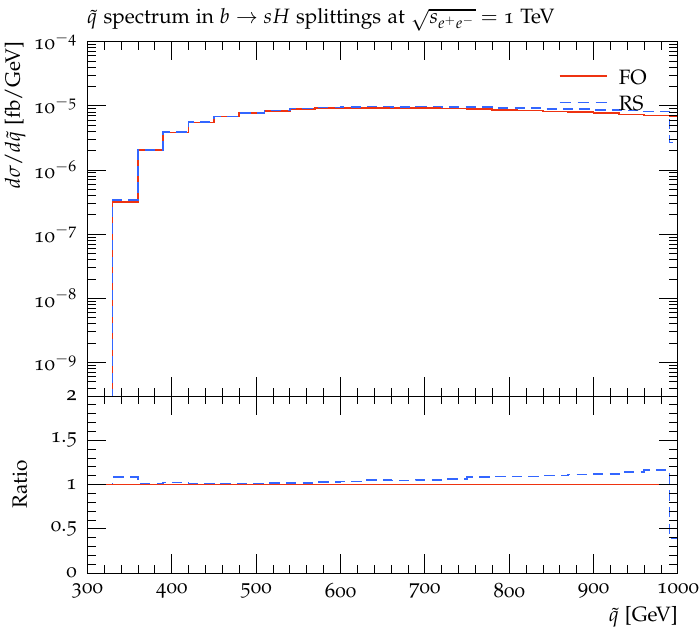}
        \caption{}
        \label{f:bsh2 mh2-130 qt}
    \end{subfigure}
    \begin{subfigure}[b]{0.49\textwidth}
        \centering
        \includegraphics[width=\textwidth]{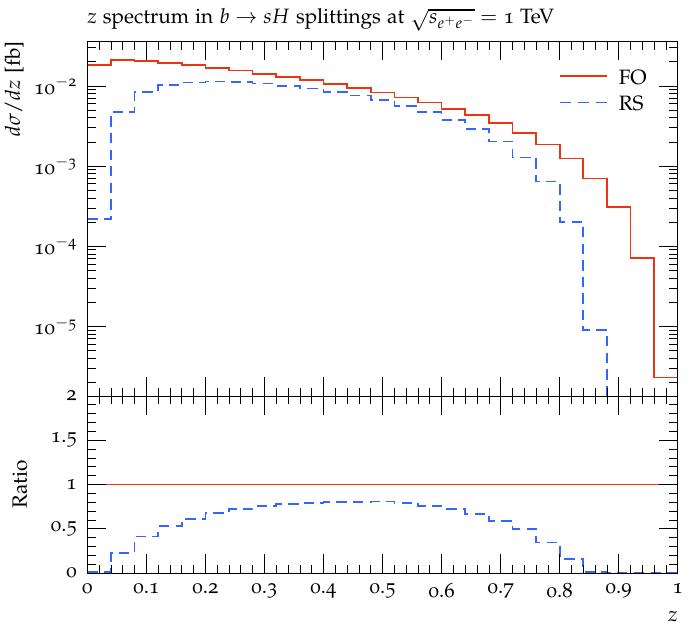}
        \caption{}
        \label{f:bsh2 mh2-130 z}
    \end{subfigure}
    \begin{subfigure}[b]{0.49\textwidth}
        \centering
        \includegraphics[width=\textwidth]{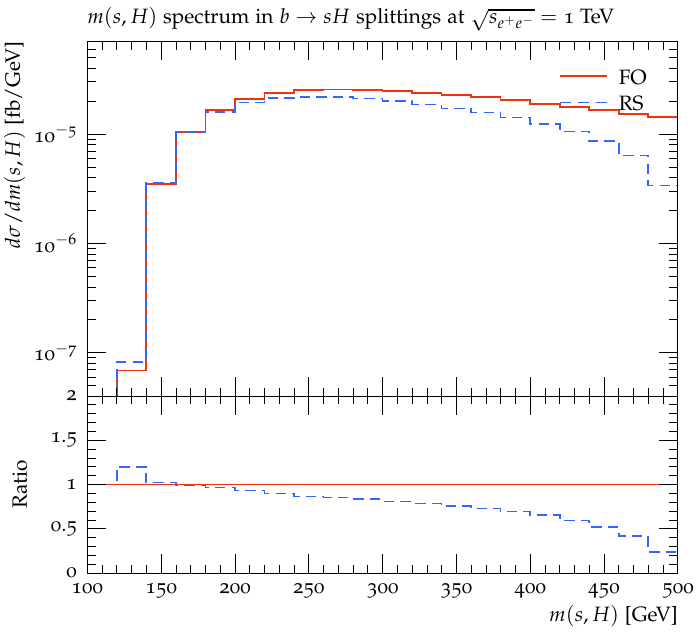}
        \caption{}
        \label{f:bsh2 mh2-130 drri}
    \end{subfigure}
    \begin{subfigure}[b]{0.49\textwidth}
        \centering
        \includegraphics[width=\textwidth]{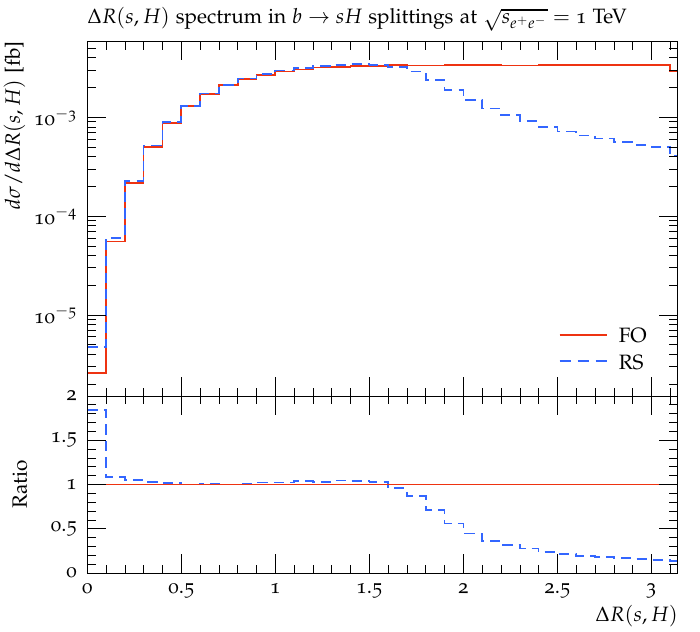}
        \caption{}
        \label{f:bsh2 mh2-130 mij}
    \end{subfigure}
    \caption{Performance test for $b(\Bar{b})\rightarrow s(\Bar{s}) H$ branching in \textsf{Herwig~7}, where the mass of the $H$ boson is assumed as 130 GeV. The notation of the figure is the same as in figure~\ref{f:bbh2 mh2-10}.}
    \label{f:bsh2 mh2-130}
\end{figure} 

\begin{figure} [h!]
\centering
    \begin{subfigure}[b]{0.49\textwidth}
        \centering
        \includegraphics[width=\textwidth]{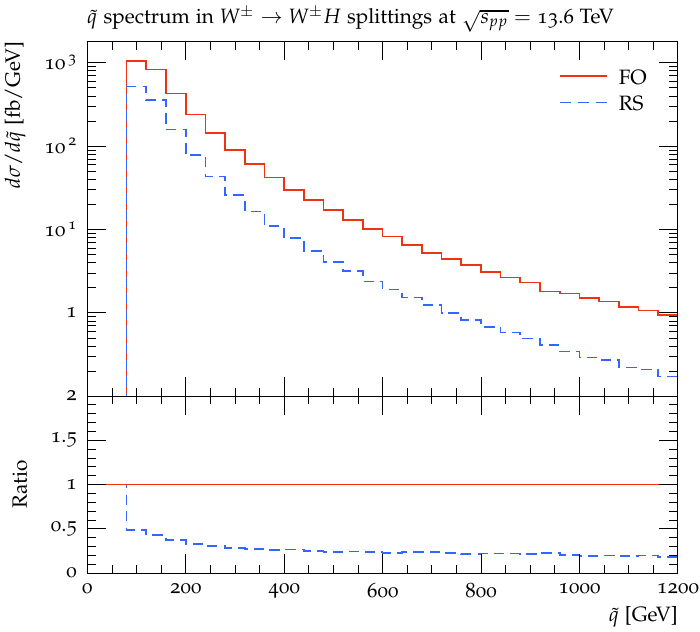}
        \caption{}
        \label{f:wh2j qt}
    \end{subfigure}
    \begin{subfigure}[b]{0.49\textwidth}
        \centering
        \includegraphics[width=\textwidth]{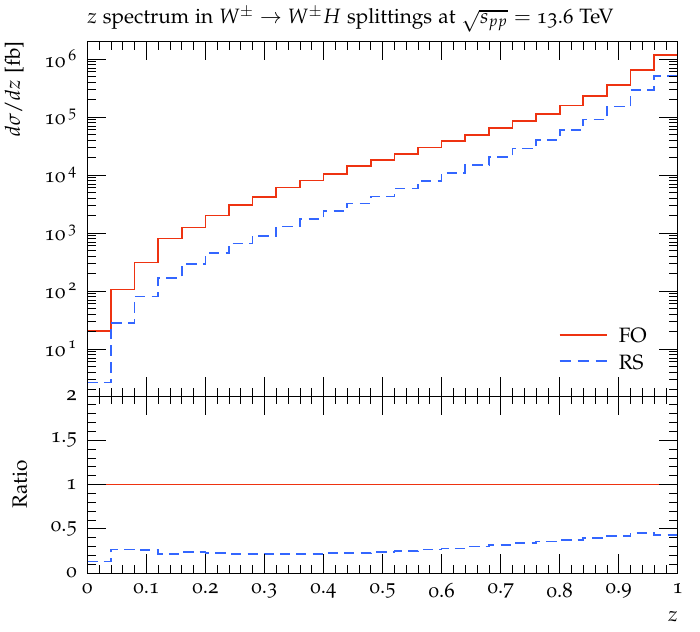}
        \caption{}
        \label{f:wh2j z}
    \end{subfigure}
    \begin{subfigure}[b]{0.49\textwidth}
        \centering
        \includegraphics[width=\textwidth]{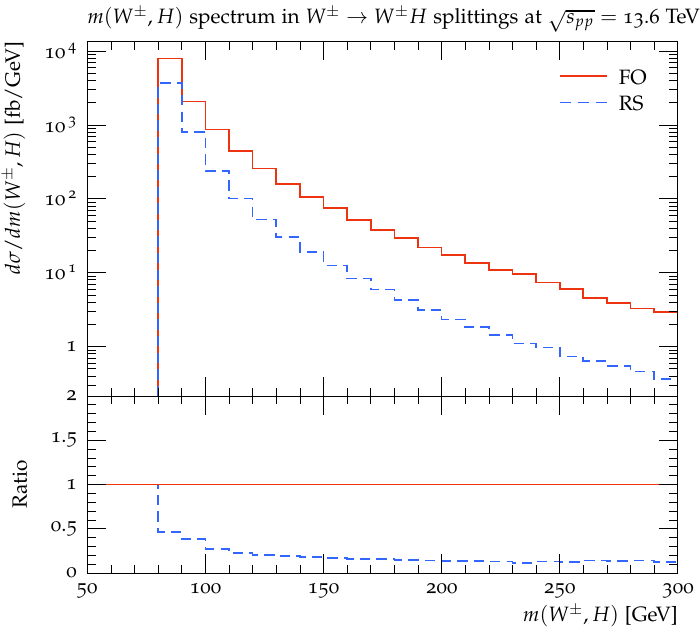}
        \caption{}
        \label{f:wh2j drri}
    \end{subfigure}
    \begin{subfigure}[b]{0.49\textwidth}
        \centering
        \includegraphics[width=\textwidth]{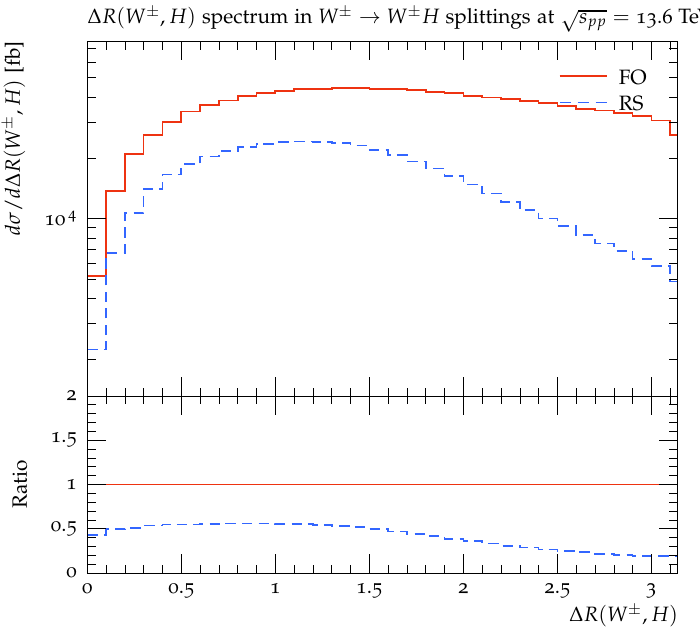}
        \caption{}
        \label{f:wh2j mij}
    \end{subfigure}
    \caption{Performance test for $W^\pm \rightarrow W^\pm H$ branching at the $pp\rightarrow W^\pm j$ process with the centre-of-mass energy of 13.6 TeV in \textsf{Herwig~7}, where the mass of the $H$ boson is assumed as 1 GeV. The notation of the figure is the same as in figure~\ref{f:bbh2 mh2-10}.}
    \label{f:wh2j}
\end{figure}

\begin{figure} [h!]
\centering
    \begin{subfigure}[b]{0.49\textwidth}
        \centering
        \includegraphics[width=\textwidth]{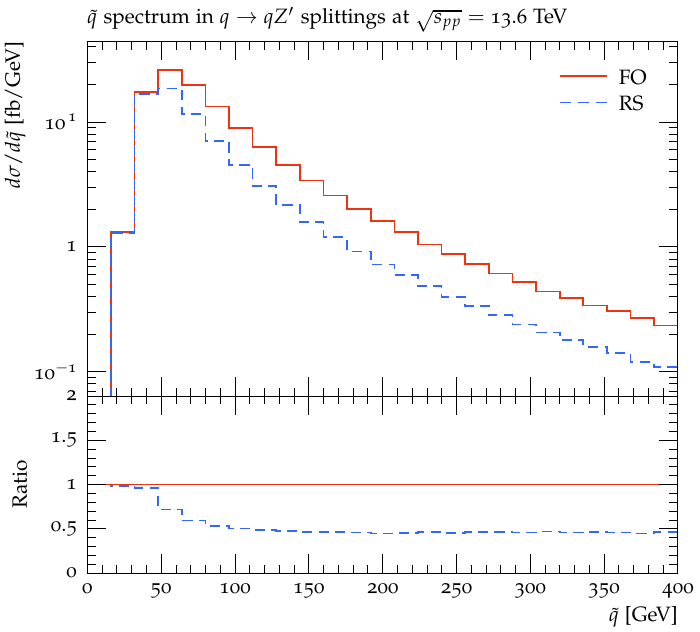}
        \caption{}
        \label{f:ddzp qt}
    \end{subfigure}
    \begin{subfigure}[b]{0.49\textwidth}
        \centering
        \includegraphics[width=\textwidth]{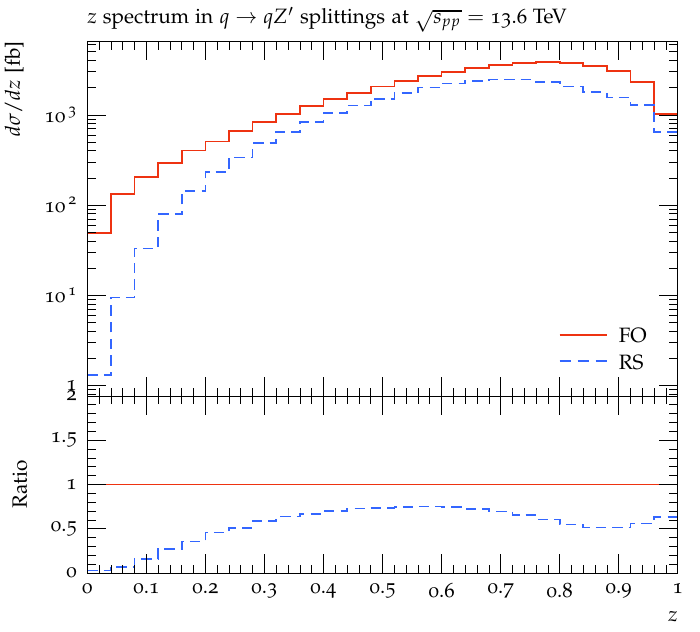}
        \caption{}
        \label{f:ddzp z}
    \end{subfigure}
    \begin{subfigure}[b]{0.49\textwidth}
        \centering
        \includegraphics[width=\textwidth]{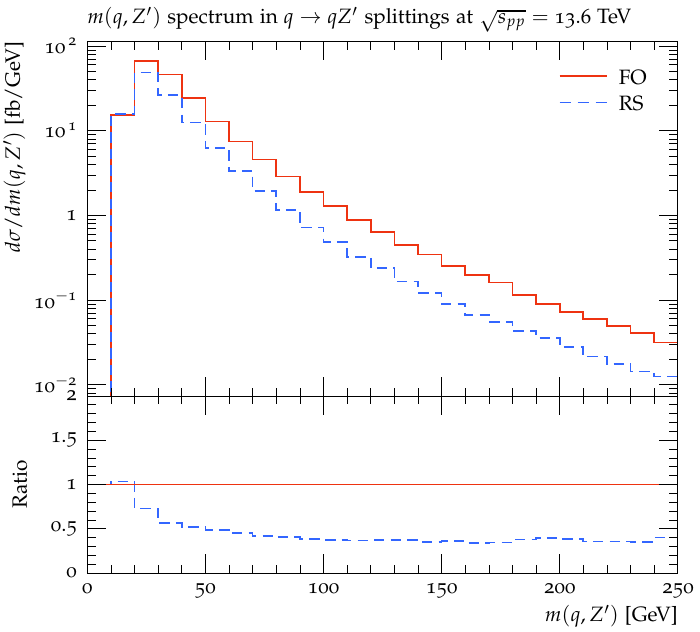}
        \caption{}
        \label{f:ddzp drri}
    \end{subfigure}
    \begin{subfigure}[b]{0.49\textwidth}
        \centering
        \includegraphics[width=\textwidth]{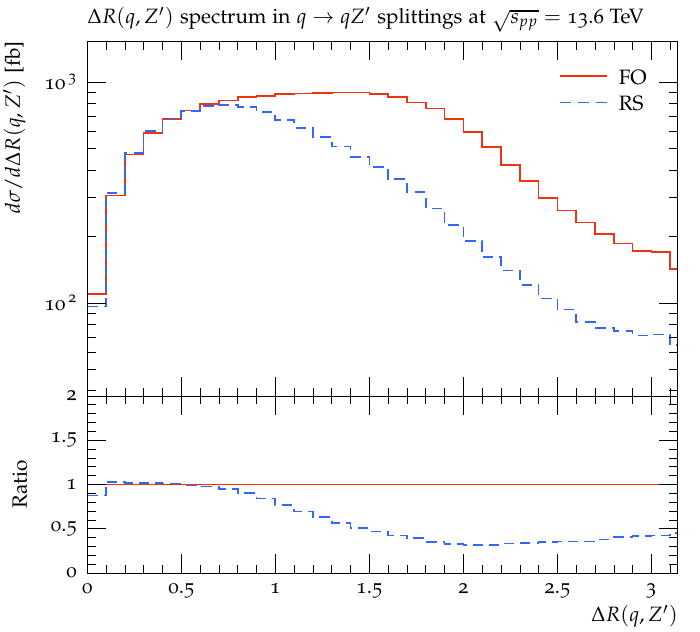}
        \caption{}
        \label{f:ddzp mij}
    \end{subfigure}
    \caption{Performance test for $q\rightarrow qZ'$ branching in \textsf{Herwig~7}, where the mass of the $Z'$ boson is assumed as 10 GeV. The notation of the figure is the same as in figure~\ref{f:bbh2 mh2-10}.}
    \label{f:ddzp}
\end{figure} 

\begin{figure} [h!]
\centering
    \begin{subfigure}[b]{0.49\textwidth}
        \centering
        \includegraphics[width=\textwidth]{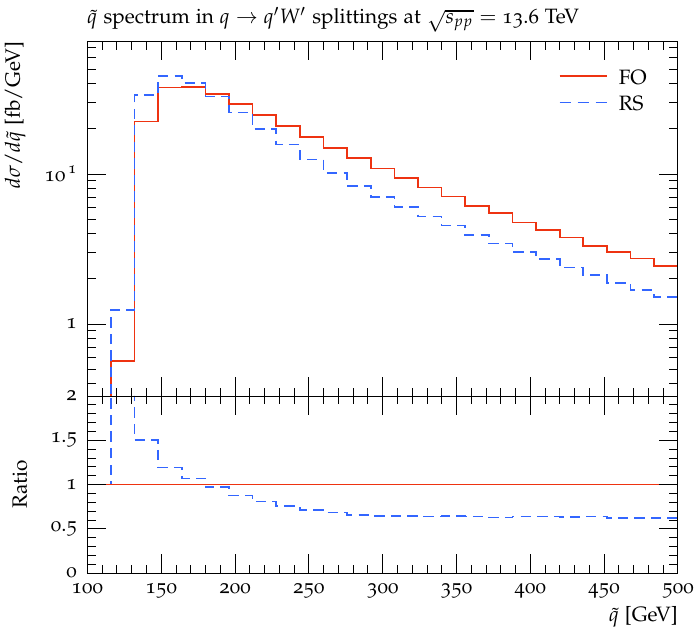}
        \caption{}
        \label{f:cswp qt}
    \end{subfigure}
    \begin{subfigure}[b]{0.49\textwidth}
        \centering
        \includegraphics[width=\textwidth]{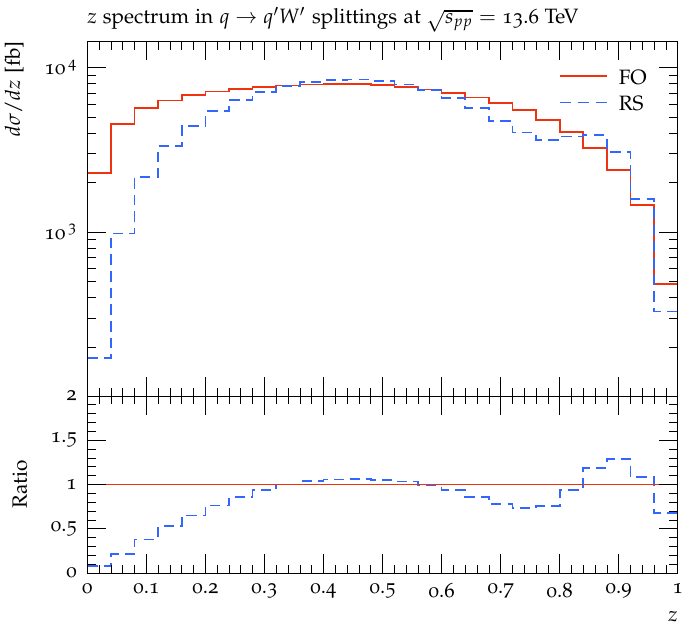}
        \caption{}
        \label{f:cswp z}
    \end{subfigure}
    \begin{subfigure}[b]{0.49\textwidth}
        \centering
        \includegraphics[width=\textwidth]{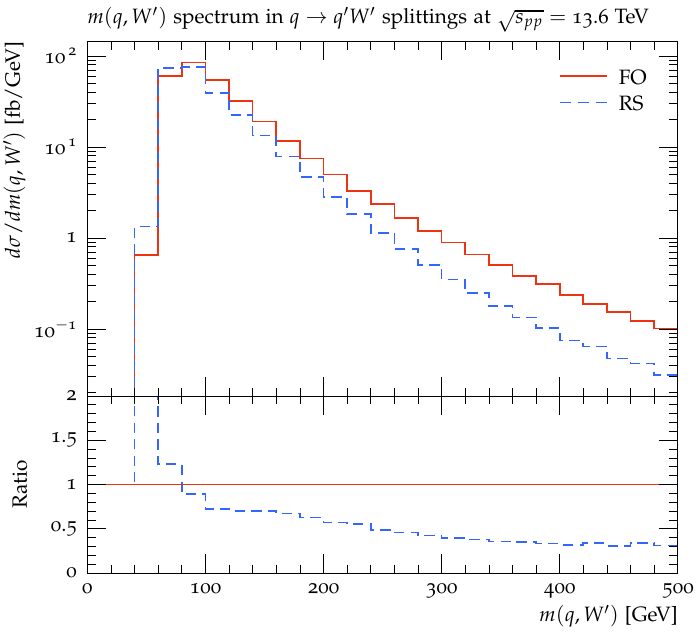}
        \caption{}
        \label{f:cswp drri}
    \end{subfigure}
    \begin{subfigure}[b]{0.49\textwidth}
        \centering
        \includegraphics[width=\textwidth]{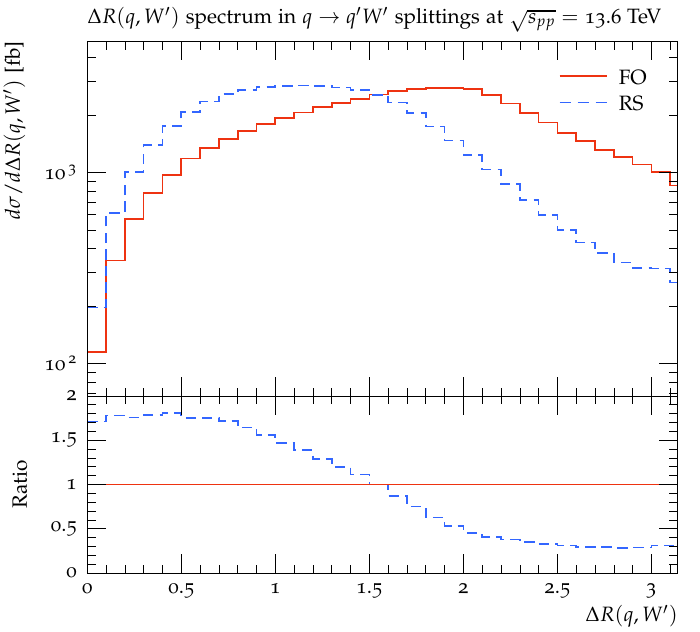}
        \caption{}
        \label{f:cswp mij}
    \end{subfigure}
    \caption{Performance test for $q\rightarrow qW'$ branching in \textsf{Herwig~7}, where the mass of the $W'$ boson is assumed as 50 GeV. The notation of the figure is the same as in figure~\ref{f:bbh2 mh2-10}.}
    \label{f:cswp}
\end{figure}

\begin{figure} [h!]
\centering
    \begin{subfigure}{0.49\textwidth}
    \centering
    \includegraphics[width=\textwidth]{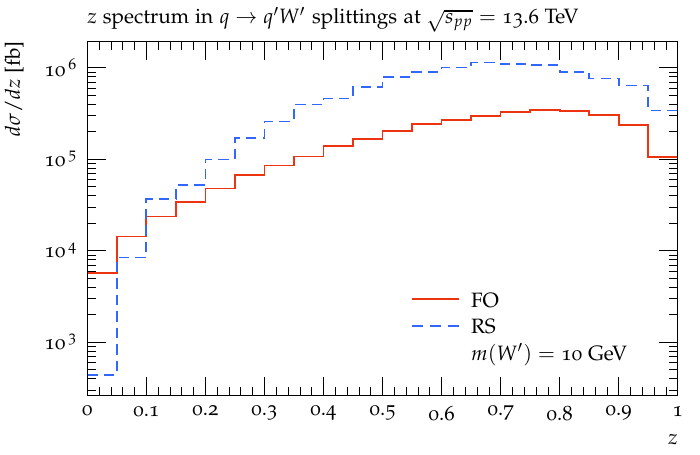}
    \caption{}
    \label{f:wp10}
    \end{subfigure}
    \begin{subfigure}{0.49\textwidth}
    \centering
    \includegraphics[width=\textwidth]{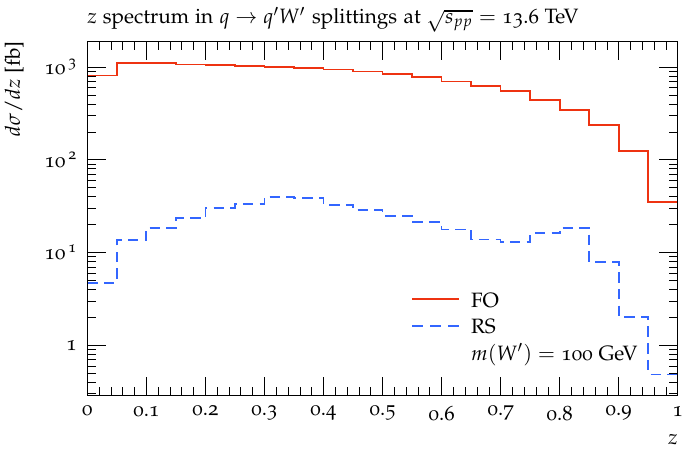}
    \caption{}
    \label{f:wp100}
    \end{subfigure}
    \begin{subfigure}{0.49\textwidth}
    \centering
    \includegraphics[width=\textwidth]{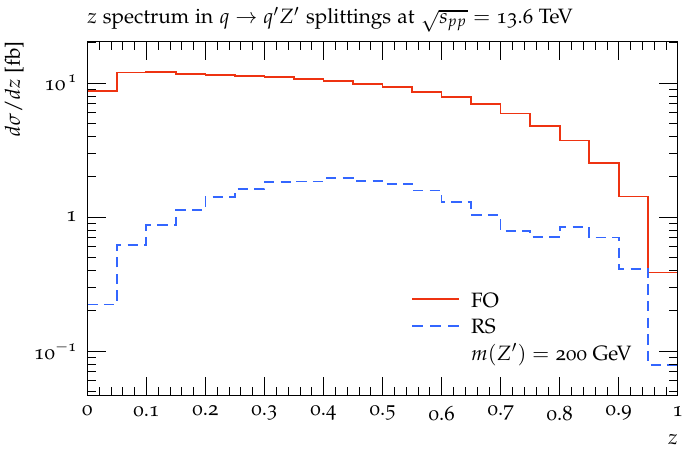}
    \caption{}
    \label{f:zp200}
    \end{subfigure}
    \caption{$z$ distributions for $q\rightarrow q' V'$ branching in \textsf{Herwig~7} at the $\sqrt{s} = 13.6$ TeV proton-proton collision. (a) $q\rightarrow q' W'$ branching with $m(W') =$ 10 GeV. (b) $q\rightarrow q' W'$ branching with $m(W') =$ 100 GeV. (c) $q\rightarrow q' Z'$ branching with $m(Z') =$ 200 GeV.}
    \label{f:qqv}
\end{figure} 

\end{document}